\documentclass[aps,twocolumn,10pt,superscriptaddress,prb]{revtex4-1}

\pdfoutput=1

\usepackage{mathtools}
\usepackage{subfigure}
\usepackage{amsmath}
\usepackage{amssymb}
\usepackage{physics}
\usepackage{hyperref}
\usepackage{dsfont}
\usepackage{leftidx}
\usepackage{chngcntr}
\usepackage{color}

\DeclareMathAlphabet{\pazocal}{OMS}{zplm}{m}{n}

\definecolor{emerald}{rgb}{0.31, 0.78, 0.47}

\hypersetup{
     colorlinks=true,
     linkcolor=blue,
     filecolor=blue,
     citecolor=emerald,      
     urlcolor=magenta,
}

\newcommand{\f}[2]{\frac{#1}{#2}} 

\newcommand{\bo}[1]{\boldsymbol{#1}}

\newcommand{\D}{\mathrm{d}}

\makeatletter
\newcommand*{\rom}[1]{\expandafter\@slowromancap\romannumeral #1@}
\makeatother

\makeatletter
\newsavebox{\@brx}
\newcommand{\llangle}[1][]{\savebox{\@brx}{\(\m@th{#1\langle}\)}%
  \mathopen{\copy\@brx\kern-0.5\wd\@brx\usebox{\@brx}}}
\newcommand{\rrangle}[1][]{\savebox{\@brx}{\(\m@th{#1\rangle}\)}%
  \mathclose{\copy\@brx\kern-0.5\wd\@brx\usebox{\@brx}}}
\makeatother

\begin{document}

\title{Superconducting order of $\mathrm{Sr}_2\mathrm{RuO}_4$ from a three-dimensional microscopic model}

\author{Henrik S.~R{\o}ising}
\email[]{henrik.roising@physics.ox.ac.uk}
\affiliation{Rudolf Peierls Center for Theoretical Physics, Oxford OX1 3PU, United Kingdom}

\author{Thomas Scaffidi}
\affiliation{Department of Physics, University of California, Berkeley, CA 94720, USA}
\affiliation{Department of Physics, University of Toronto, Toronto, Ontario, M5S 1A7, Canada}

\author{Felix Flicker}
\affiliation{Rudolf Peierls Center for Theoretical Physics, Oxford OX1 3PU, United Kingdom}

\author{Gunnar F.~Lange}
\affiliation{Rudolf Peierls Center for Theoretical Physics, Oxford OX1 3PU, United Kingdom}
\affiliation{TCM Group, Cavendish Laboratory, University of Cambridge, Cambridge CB3 0HE, United Kingdom}

\author{Steven H.~Simon}
\affiliation{Rudolf Peierls Center for Theoretical Physics, Oxford OX1 3PU, United Kingdom}

\date{\today}

\begin{abstract}
We compute and compare even- and odd-parity superconducting order parameters of strontium ruthenate ($\mathrm{Sr}_2\mathrm{RuO}_4$) in the limit of weak interactions, resulting from a fully microscopic three-dimensional model including spin-orbit coupling. We find that odd-parity helical and even-parity $d$-wave order are favored for smaller and larger values of the Hund's coupling parameter $J$, respectively. Both orders are found compatible with specific heat data and the recently-reported nuclear magnetic resonance (NMR) Knight shift drop [A.~Pustogow \emph{et al}.~Nature \textbf{574}, 72 (2019)]. The chiral $p$-wave order, numerically very competitive with helical order, sharply conflicts with the NMR experiment.
\end{abstract}

\maketitle

\section{Introduction}
\label{sec:Introduction}
Superconductivity was discovered in the layered perovskite strontium ruthenate, Sr$_2$RuO$_4$ (SRO), about $25$ years ago~\cite{MaenoEA94}. Muon spin relaxation and Kerr effect experiments indicated time-reversal symmetry breaking (TRSB) in the superconducting phase~\cite{LukeEA98, XiaEA06}. The accompanied absence of a drop in the spin susceptibility~\cite{IshidaEA98} pointed toward a chiral $p$-wave order parameter~\cite{MackenzieMaeno03, MaenoEA12, Kallin16}, which would make SRO an electronic analog of the A-phase of ${}^3\mathrm{He}$~\cite{Leggett75, RiceSigrist95}. In addition to the general interest in instances of unconventional superconductivity, intrinsic chiral $p$-wave superconductors are of particular importance owing to the possibility of enabling topological quantum computation with non-Abelian anyons~\cite{ReadGreen2000, Ivanov01, NayakEA08}.

However, a series of key experiments conflict with the above interpretation. The linear temperature dependence of specific heat~\cite{NishiZakiEA00} at low temperature implies nodes or deep minima in the gap~\cite{DeguchiEA04}. Recent thermal Hall conductivity measurements further suggest vertical (out-of-plane) line nodes~\cite{HassingerEA17, DodaroEA18}. Uniaxial strain experiments see no indications of a $T_c$-cusp, as expected for chiral $p$-wave order~\cite{HicksEA14, SteppkeEA17, LiEA19}. A very recent in-plane field NMR experiment measured a significant spin susceptibility drop~\cite{PustogowEA19}, contradicting the original measurements, which in the absence of spin-orbit coupling (SOC) would exclude all models featuring vectorial order parameters (triplet) pointing out of the basal plane~\cite{BalianWerthamer63}. This has reignited a longstanding debate, possibly making the case of helical or even-parity order parameters plausible~\cite{RomerEA19, RamiresSigrist19, LiEA19, TankaEA09}. We note, however, that strong SOC~\cite{Damascelli08, DamascelliEA14, TamaiEA18} in a multi-orbital system complicates the analysis of the magnetic susceptibility compared to the single orbital case~\cite{RiceSigrist95}.

Most studies so far have used a two-dimensional (2D) model, taking advantage of the quasi-2D nature of the dispersion relation of the relevant bands. However, the small corrugation of the cylindrical Fermi surfaces is deceptive, and actually hides a non-trivial $k_z$ dependence of the orbital content of the bands due to SOC, whose effect was shown to be three-dimensional (3D)~\cite{Damascelli08}. A full 3D calculation is therefore warranted in order to study superconductivity in SRO~\cite{Damascelli08, MackenzieEA17, HassingerEA17, RoisingEA18, RamiresSigrist19, HuangNematicEA19, GingrasEA19}, and especially to study the effect of SOC on the recent Knight shift experiments. Further, a 3D calculation is also required to study the possibility of horizontal line nodes, which have been proposed as a way to reconcile a nodal superconducting gap with TRSB~\cite{ZuticMazin05}.

We propose an effective three-band three-dimensional model with on-site interaction, and we calculate the superconducting order parameter in the weak-coupling limit. As a function of the ratio of the Hund's coupling $J$ to the Hubbard interaction strength $U$ we find a transition at $J/U \approx 0.15$ from an odd-parity helical phase with accidental (near-)nodes to an even-parity phase with symmetry-imposed vertical line nodes, both of which are compatible with several key experiments, but are not compatible with the observation of TRSB. This will be commented on in the Conclusions.

The paper is organized as follows. In Section \ref{sec:Model} we describe the effective three-dimensional tight-binding model considered in this work. In Section \ref{sec:Method} we outline the weak-coupling approach used to numerically determine the superconducting order parameter. The main results are presented in Section \ref{sec:Results}, where we provide detailed calculations of -- and compare results with -- two key experimental probes: the specific heat (Section \ref{sec:SpecificHeat}) and the NMR Knight shift (Section \ref{sec:MagneticSusceptibility}). We conclude and discuss future prospects in Section \ref{sec:Conclusions}.

\section{Effective three-dimensional model}
\label{sec:Model}
In SRO, three bands cross the Fermi energy and form quasi-two-dimensional Fermi surfaces, commonly denoted $\alpha$, $\beta$, and $\gamma$~\cite{DamascelliEA00, BergmannEA00, BergmannEA03, TamaiEA18}. Sheets $\alpha$ and $\beta$ are formed mostly by the $4d_{xz}$ and $4d_{yz}$ ruthenium (Ru) orbitals, whereas the $\gamma$ sheet stems mostly from the $4d_{xy}$ Ru orbital. We construct a tight-binding model for the three active bands, based on the three Ru $t_{2g}$ orbitals:
\begin{equation}
H_0 = \sum_{\bo{k}, s} \bo{\psi}^{\dagger}_s(\bo{k}) \pazocal{H}_s(\bo{k}) \bo{\psi}_s(\bo{k})
\label{eq:Hamiltonian}
\end{equation}
where
\begin{equation}
\pazocal{H}_s(\bo{k}) = \begin{pmatrix}
\varepsilon_{AA}(\bo{k}) & \varepsilon_{AB}(\bo{k}) - i s\eta & \varepsilon_{AC}(\bo{k}) + i\eta \\
\varepsilon_{BA}(\bo{k}) + i s\eta & \varepsilon_{BB}(\bo{k}) & \varepsilon_{BC}(\bo{k}) - s\eta \\ \varepsilon_{CA}(\bo{k}) - i\eta & \varepsilon_{CB}(\bo{k}) -s\eta & \varepsilon_{CC}(\bo{k})
\end{pmatrix},
\label{eq:KineticHamiltonianMain}
\end{equation}
and $\bo{\psi}_s(\bo{k}) = [c_{A s}(\bo{k}), \hspace{1mm} c_{B s}(\bo{k}), \hspace{1mm} c_{C \bar{s}}(\bo{k})]^T$. We here used the shorthand notation $A=xz$, $B=yz$, $C=xy$, $\bar{s} = -s$, and $s$ denotes spin ($s=+1$ is understood as $\uparrow$ and $s = -1$ means $\downarrow$). The annihilation operator for an electron with wavevector $\mathbf{k}$ and spin $s$ on Ru orbital $4d_a$ is denoted by $c_{a s}\left(\mathbf{k}\right)$. The matrix elements $\varepsilon_{ab}(\bo{k})$ account for intra- and inter-orbital hopping, both in- and out-of-plane, and $\eta$ sets the SOC amplitude. \emph{A priori} we retain terms up to three sites apart in-plane and leading order terms, including inter-orbital terms, out-of-plane:
\begin{widetext}
\begin{align}
\varepsilon_{\mathrm{1D}}(k_{\parallel}, k_{\perp}, k_z) &= - 2t_1 \cos(k_{\parallel}) - 2t_2 \cos(k_{\perp}) - 2t_3 \cos(2k_{\parallel}) - 4t_4 \cos(k_{\parallel}) \cos(k_{\perp}) - 4t_5 \cos(2 k_{\parallel}) \cos(k_{\perp}) \nonumber \\
& \hspace{15pt} - 2t_6 \cos(3k_{\parallel}) - 2t_7\cos(2k_{\perp}) - 2t_8 \cos(k_{\parallel}/2) \cos(k_{\perp}/2) \cos(k_z/2)  - \mu_{\mathrm{1D}},  \label{eq:1DHopping} \\
\varepsilon_{\mathrm{2D}}(\bo{k}) &= - 2\bar{t}_1 \left[ \cos(k_x) + \cos(k_y) \right] - 4\bar{t}_2 \cos(k_x) \cos(k_y) - 2\bar{t}_3 \left[ \cos(2 k_x) + \cos(2 k_y) \right] \nonumber \\
& \hspace{15pt} - 4\bar{t}_{4} \left[ \cos(2 k_x) \cos(k_y) + \cos(2 k_y) \cos(k_x) \right]  - 2\bar{t}_{5} \cos(k_z/2) \cos(k_x/2) \cos(k_y/2) - \mu_{\mathrm{2D}}, \label{eq:2DHopping} \\
\varepsilon_{AB}(\bo{k}) &= - 4t_{\mathrm{int},1} \sin(k_x)\sin(k_y) - 4t_{\mathrm{int},2} \sin(k_x/2)\sin(k_y/2) \cos(k_z/2), \label{eq:offHopping1} \\
\varepsilon_{AC}(\bo{k}) &= - 4t_{\mathrm{int},3} \sin(k_z/2)\cos(k_x/2) \sin(k_y/2), \label{eq:offHopping2}  \\
\varepsilon_{BC}(\bo{k}) &= - 4t_{\mathrm{int},3} \sin(k_z/2)\sin(k_x/2) \cos(k_y/2). \label{eq:offHopping3}
\end{align}
\end{widetext}
To relate the above terms to Eq.~\eqref{eq:KineticHamiltonianMain} we set $\varepsilon_{AA}(\bo{k}) = \varepsilon_{\mathrm{1D}}(k_x, k_y, k_z)$, $\varepsilon_{BB}(\bo{k}) = \varepsilon_{\mathrm{1D}}(k_y, k_x, k_z)$, and $\varepsilon_{CC}(\bo{k}) = \varepsilon_{\mathrm{2D}}(k_x, k_y, k_z)$. All terms in this model respect the crystal symmetries; they preserve inversion and time-reversal symmetry~\cite{RamiresSigrist19}. With the above conventions the first Brillouin zone is here defined as $\text{BZ} = [-\pi,\pi]^2\times [-2\pi,2\pi]$. 

\subsection{Band structure fit with Monte Carlo sampling}
Within the $19$-dimensional parameter space 
\begin{equation}
\lbrace t \rbrace =  \big\lbrace \lbrace t_i \rbrace_{i=1}^{8}, \lbrace \bar{t}_i \rbrace_{i=1}^{5}, \lbrace t_{\mathrm{int},i} \rbrace_{i=1}^3, \mu_{\mathrm{1D}}, \mu_{\mathrm{2D}}, \eta \big\rbrace, 
\label{eq:Parameters}
\end{equation}
we seek the set $\lbrace t \rbrace$ that globally minimizes the quantity
\begin{equation}
\begin{aligned}
D&(\lbrace t \rbrace) = \sum_{\mu} \sum_{\bo{k}} w_{\bo{k}} \big( \xi_{\mu}(\bo{k}, \lbrace t \rbrace) - \tilde{\xi}_{\mu}(\bo{k}) \big)^2  \\
+& \sum_{\mu} \sum_{\bo{q}_{\mu} \in S_{\mu}} \tilde{w}_{\bo{q}_{\mu}} \big( \lvert u^{\mu}_{C}(\bo{q}_{\mu}, \lbrace t \rbrace) \rvert^2 - \lvert \tilde{u}^{\mu}_{C}(\bo{q}_{\mu}) \rvert^2 \big)^2,
\label{eq:QantityD}
\end{aligned}
\end{equation}
where $\mu = \alpha, \beta, \gamma$ is the band index, $\xi_{\mu}$ ($u_{C}^{\mu}$) is the band energy (orbital content $C=xy$, as given by the eigenvector components of $\pazocal{H}_{s=+1}$) of the model in Eq.~\eqref{eq:KineticHamiltonianMain}. Similarly, $\tilde{\xi}_{\mu}$ ($\tilde{u}^{\mu}_{C}$) is the band energy (orbital content) of the model of Ref.~\onlinecite{DamascelliEA14}, which is based on a tight-binding fit from spin-resolved angle-resolved photoemission spectroscopy (ARPES) data. Finally, $w_{\bo{k}}$ ($\tilde{w}_{\bo{k}}$) are chosen energy (orbital) weights. For the $\bo{q}_{\mu}$'s we choose the three in-plane directions $\theta = 0, \hspace{1mm} \pi/6, \hspace{1mm} \pi/4$ for the three $k_z$ cuts $0, \hspace{1mm} \pi, \hspace{1mm} 2\pi$.

\begin{figure}[h!tb]
	\centering
	\includegraphics[width=0.90\linewidth]{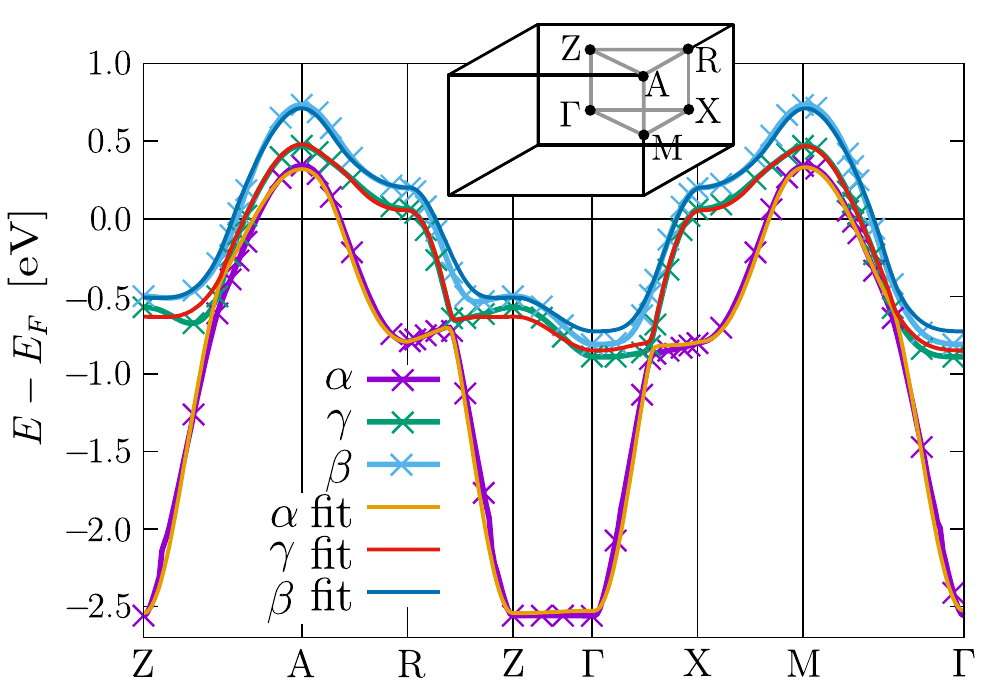} 
	\caption{Tight-binding fit to the 17-band model of Ref.~\onlinecite{DamascelliEA14}. The inset indicates paths and high-symmetry points in the Brillouin zone, using a primitive tetragonal unit cell. The crosses mark the chosen fitting points, i.e.~the $\bo{k}$-path.}
	\label{fig:FitSample}
\end{figure} 
\begin{figure*}[h!tb]
	\centering
	\subfigure[]{\includegraphics[width=0.31\linewidth]{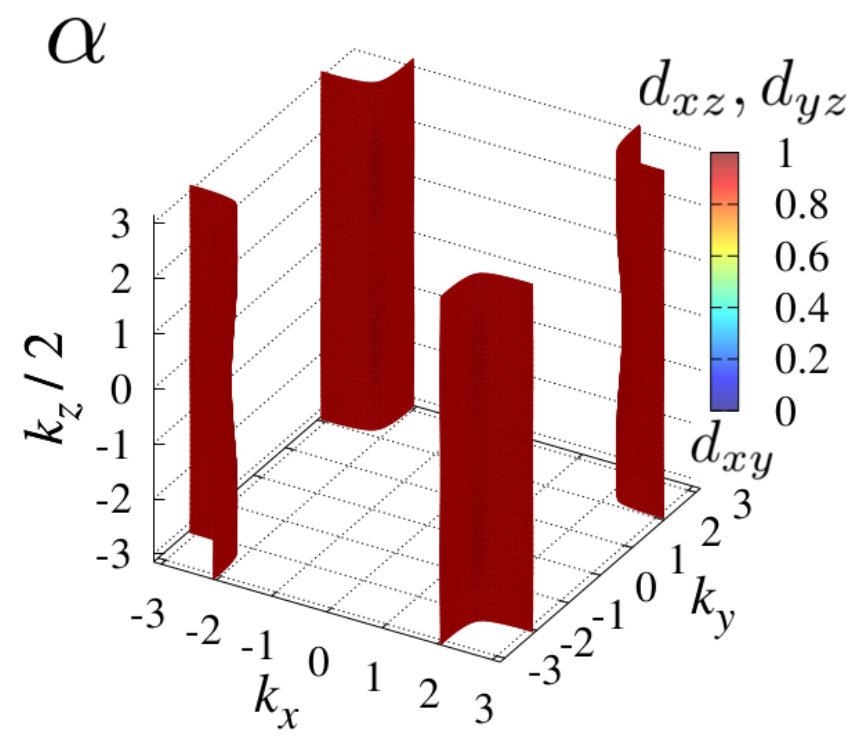}} \quad 
	\subfigure[]{\includegraphics[width=0.31\linewidth]{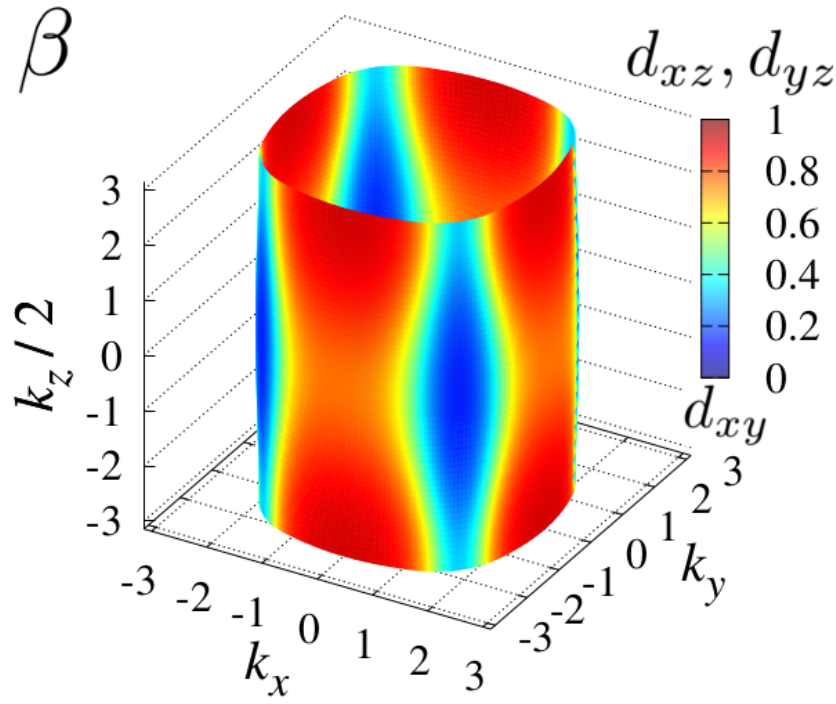}} \quad 
	\subfigure[]{\includegraphics[width=0.31\linewidth]{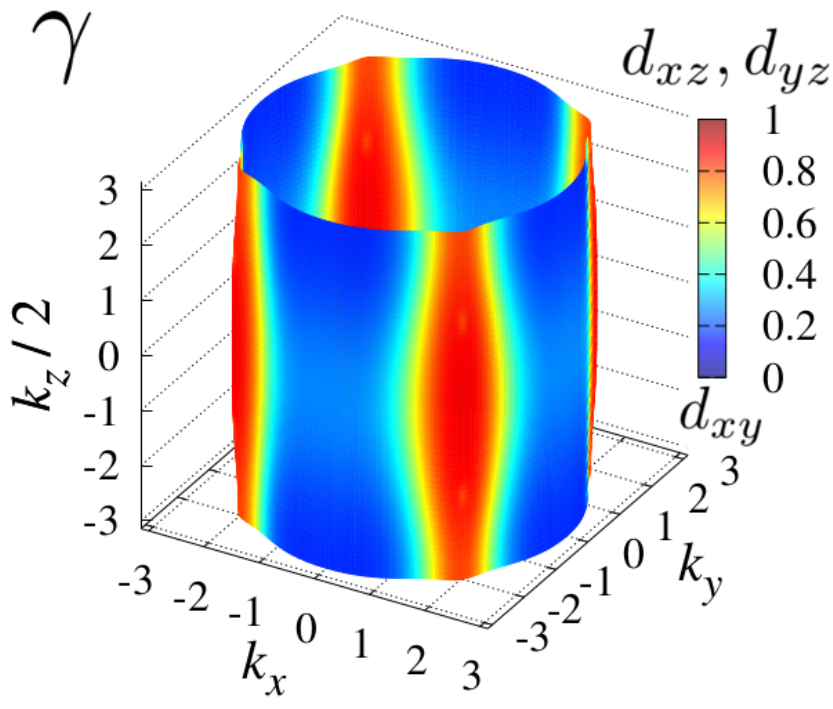}} \quad 
	\subfigure[]{\includegraphics[width=0.25\linewidth]{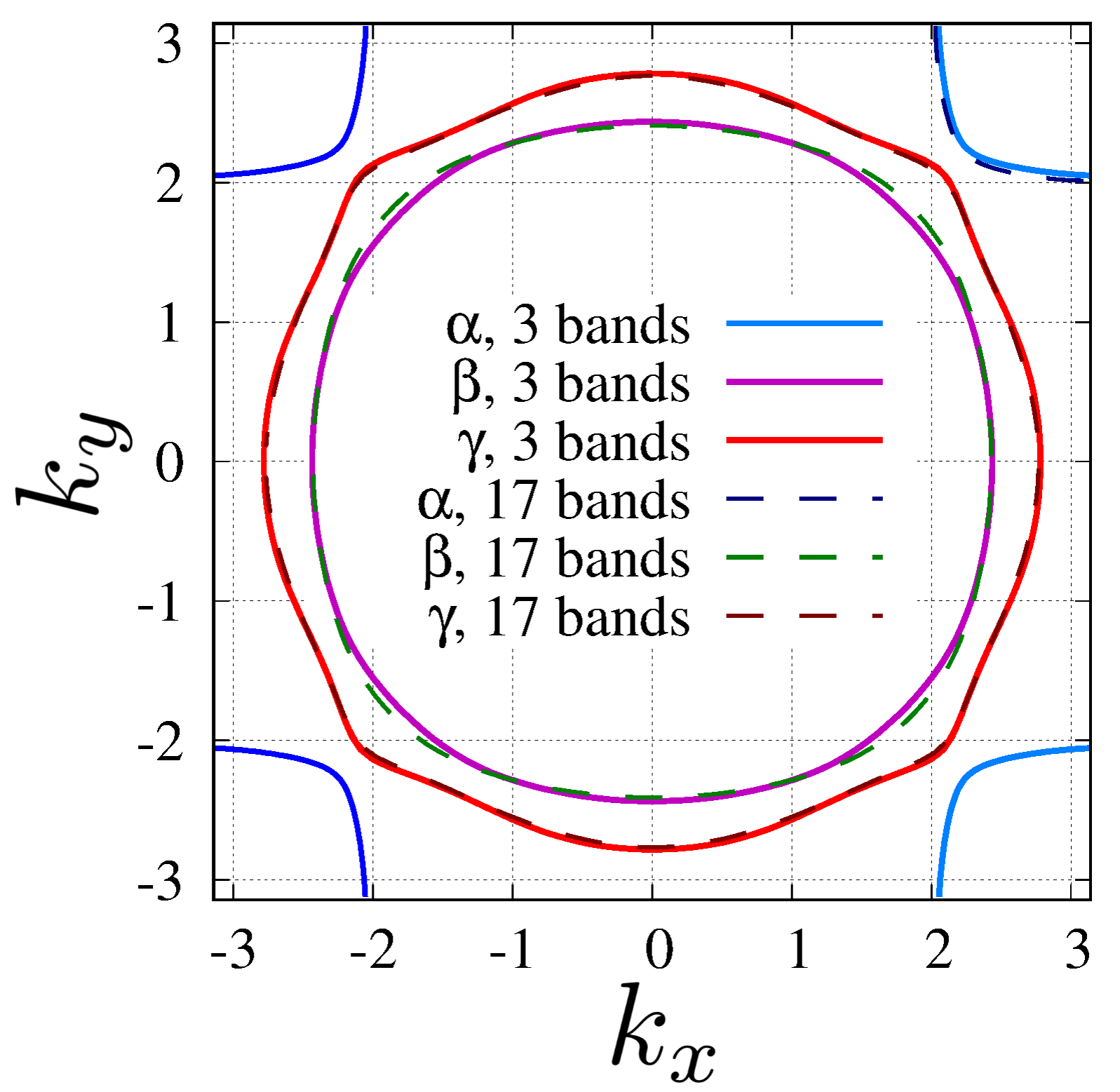}} \quad 
	\subfigure[]{\includegraphics[width=0.25\linewidth]{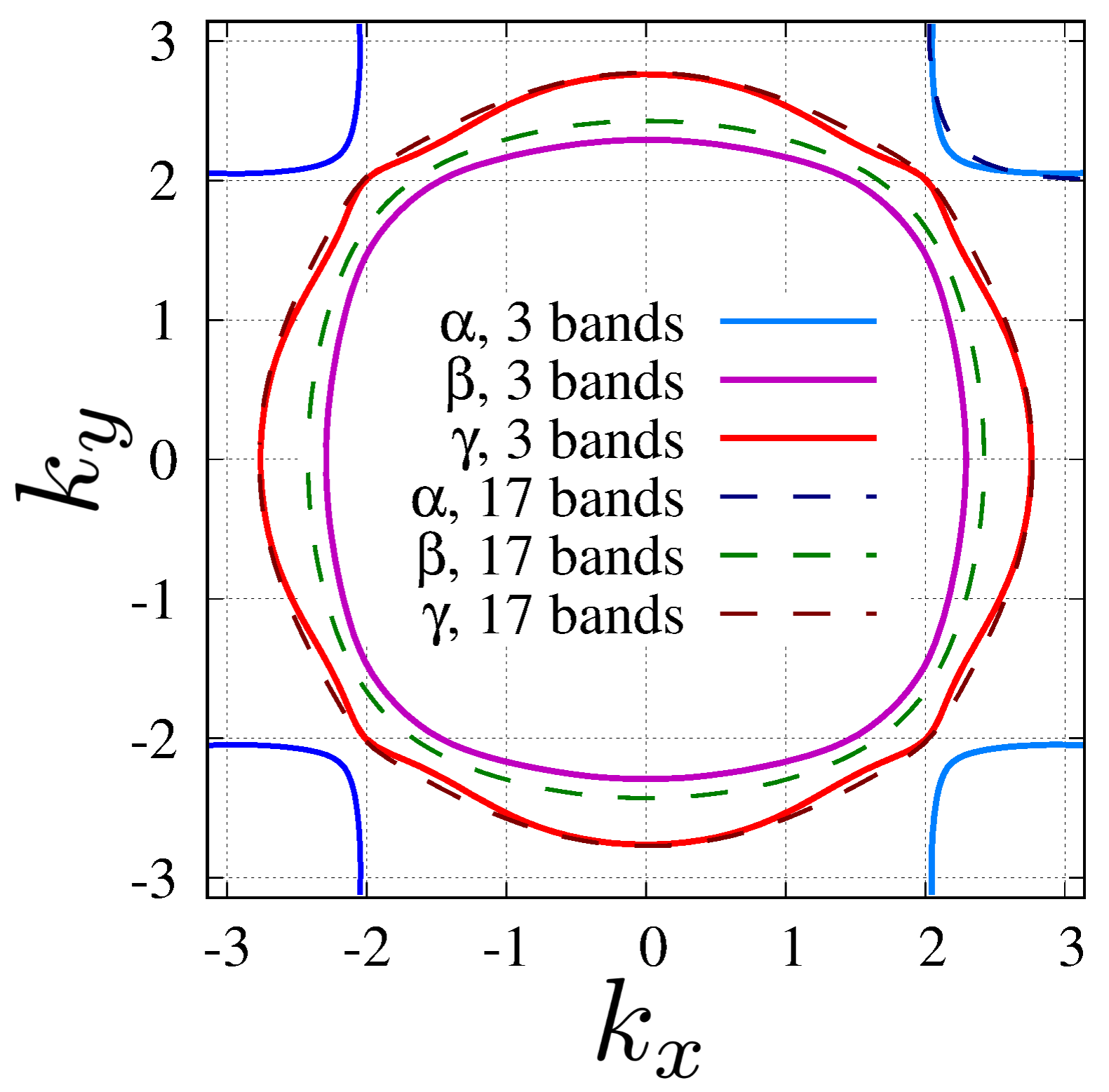}} \quad 
	\subfigure[]{\includegraphics[width=0.35\linewidth]{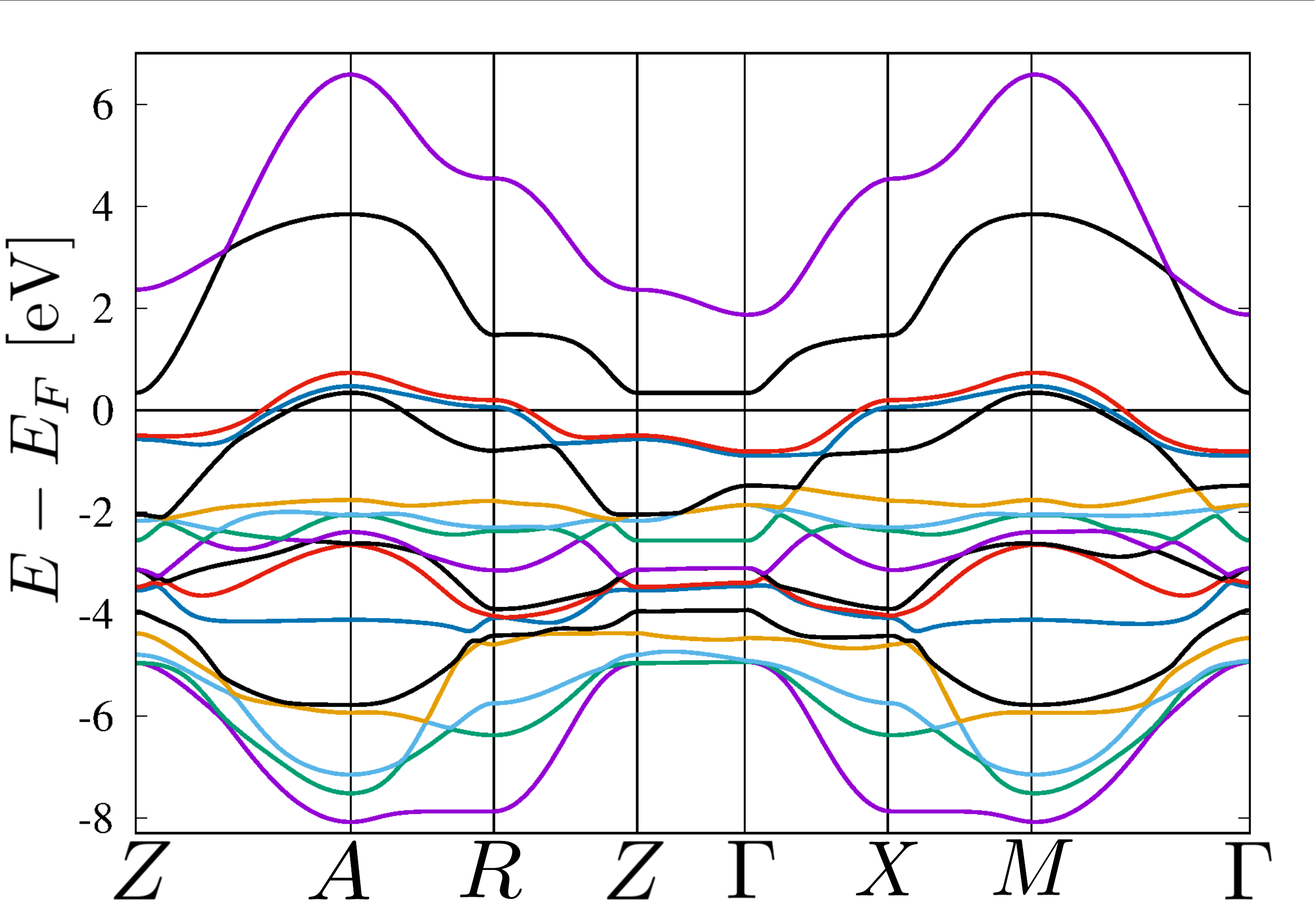}}
	\caption{(a) -- (c) Fermi surfaces with the effective model, with the color referring to the orbital content $ \lvert u^{\mu}_{xz}(\bo{k}) \rvert^2 + \lvert u^{\mu}_{yz}(\bo{k}) \rvert^2 $. The $\alpha$ sheet is  dominated by the $d_{xz}$  and $d_{yz}$ orbitals, $\beta$ is mostly dominated by $d_{xz}$ and $d_{yz}$, and $\gamma$ band mostly by $d_{xy}$, but strong mixing takes place along the $\Gamma$ -- $M$ line for the two latter bands (cf.~Ref.~\onlinecite{DamascelliEA14}). Relative densities produced with this model: $\rho_{\alpha}/\rho_{\mathrm{tot}} = 0.160$, $\rho_{\beta}/\rho_{\mathrm{tot}} = 0.334$, and $\rho_{\gamma}/\rho_{\mathrm{tot}} = 0.506$ (cf.~Ref.~\onlinecite{Scaffidi2017}). (d), (e): $k_z$ slices of the Fermi surfaces with the effective model compared with the 17-band model. The cuts are taken at (d) $k_z = 0$, and (e) $k_z = 2\pi$. (f) The band structure of the 17-band model in Ref.~\onlinecite{DamascelliEA14}, which is based on spin-resolved ARPES data.}
	\label{fig:FermiSurfaces}
\end{figure*}

The hopping amplitudes were obtained by fitting the dispersion and orbital content of the 17-band model of Ref.~\onlinecite{DamascelliEA14} using Monte Carlo (MC) sampling, see Fig.~\ref{fig:FitSample}. Specifically, we draw a set $\lbrace t \rbrace$ for each MC cycle and accept it if it makes $D(\lbrace t \rbrace)$ smaller than the previously drawn set. Otherwise, it is retained as the new optimal set with probability $\exp(-D(\lbrace t \rbrace)^{1/2}/T)$, where $T$ is an artificially introduced `temperature' that we gradually lower. For the momentum path we fit the band structure at the fitting points marked with crosses in Fig.~\ref{fig:FitSample}. The points ($Z$, $\Gamma$, $M$, $X$, $A$, $R$) are weighted four times as much as the majority of the points, and points close to the Fermi energy are weighted four times as much as the remaining points. The above-mentioned points are defined in the primitive tetragonal unit cell as $\Gamma=\left(000\right)$, $Z=\left(00\frac{1}{2}\right)$, $R=\left(\frac{1}{2}0\frac{1}{2}\right)$, $X=\left(\frac{1}{2}00\right)$, $M=\left(\frac{1}{2}\frac{1}{2}0\right)$, $A=\left(\frac{1}{2}\frac{1}{2}\frac{1}{2}\right)$. The orbital weights $\tilde{w}_{\bo{q}}$ were fixed to be comparatively smaller than the energy weights $w_{\bo{k}}$.
\begin{table*}
\centering
\caption{Tight-binding parameters for $\varepsilon_{\mathrm{1D}}$ in Eq.~\eqref{eq:1DHopping} obtained with Monte Carlo sampling.}
\begin{tabular}{p{2.2cm} p{1.0cm} p{1.0cm} p{1.0cm} p{1.0cm} p{1.0cm} p{1.0cm} p{1.0cm} p{1.0cm} p{1.0cm}}  
\toprule
Parameter & $t_1$ & $t_2$ & $t_3$ & $t_4$ & $t_5$ & $t_6$ & $t_7$ & $t_8$ & $\mu_{\mathrm{1D}}$  \\ \hline
Value [meV] & $257.8$ & $27.8$ & $-35.5$ & $-22.4$ & $-4.7$ & $-2.4$ & $3.2$ & $54.5$ & $286.9$ \\ \toprule
\label{tab:HoppingParameters1}
\end{tabular}
\end{table*}
\begin{table*}
\centering
\caption{Tight-binding parameters for Eq.~\eqref{eq:2DHopping}, \eqref{eq:offHopping1}, \eqref{eq:offHopping2}, and \eqref{eq:offHopping3} obtained with Monte Carlo sampling.}
\begin{tabular}{p{2.2cm} p{1.0cm} p{1.0cm} p{1.0cm} p{1.0cm} p{1.0cm} p{1.0cm} p{1.0cm} p{1.0cm} p{1.0cm} p{1.0cm}}  
\toprule
Parameter & $\bar{t}_1$ & $\bar{t}_2$ & $\bar{t}_3$ & $\bar{t}_4$ & $\bar{t}_5$ & $\mu_{\mathrm{2D}}$ & $\eta$  & $t_{\mathrm{int},1}$ & $t_{\mathrm{int},2}$ & $t_{\mathrm{int},3}$ \\ \hline
Value [meV] & $356.8$ & $126.3$ & $17.0$ & $22.3$ & $-4.1$ & $351.9$  & $59.2$ & $2.0$ & $-15.5$ & $-5.4$ \\ \toprule
\label{tab:HoppingParameters2}
\end{tabular}
\end{table*}

The optimal tight-binding parameters are summarized in Tables \ref{tab:HoppingParameters1} and \ref{tab:HoppingParameters2}.  The expected uncertainties for the tight-binding parameters are $\pazocal{O}(10~\text{meV})$. Since the data from the 17-band model in Ref.~\onlinecite{DamascelliEA14} incorporated terms with an accuracy threshold of $10$ meV, our effective model can not hope to accurately describe terms of lower energy than this threshold. Thus, tight-binding parameters smaller than this threshold, having a negligible effect on \emph{e.g.}~the band structure, were neglected in the implementation of the method described in Section \ref{sec:Method}.

The Fermi surfaces produced with the effective tight-binding model are shown in Fig~\ref{fig:FermiSurfaces}; in Figs.~\ref{fig:FermiSurfaces} (a) -- \ref{fig:FermiSurfaces} (c) we display the orbital content of the three Fermi surfaces, revealing that the $\beta$ and $\gamma$ sheets exhibit significant mixtures of the semi-2D orbtial ($d_{xy}$) and semi-1D Ru orbitals ($d_{xz}$ and $d_{yz}$), whereas the $\alpha$ sheet is dominated by the semi-1D orbitals. Figures \ref{fig:FermiSurfaces} (d) and (e) compare two slices of the Fermi surfaces produced with the effective model with the model considered in Ref.~\onlinecite{DamascelliEA14}, and Fig.~\ref{fig:FermiSurfaces} (f) shows the full band structure of the latter model for completeness.

\section{Weak-coupling theory}
\label{sec:Method}
The projection of the Coulomb interaction on the on-site $t_{2g}$ orbitals is given by
\begin{equation}
\begin{aligned}
H_I &= \f{U}{2} \sum_{i,a,s\neq s'} n_{i a s} n_{i a s'} +  \f{U'}{2} \sum_{i, a\neq b,s, s'} n_{i a s} n_{i b s'} \\
&\hspace{10pt} + \f{J}{2} \sum_{i,a\neq b,s, s'} c_{i a s}^{\dagger}
c_{i b s'}^{\dagger}c^{\phantom{\dagger}}_{i a s'}c^{\phantom{\dagger}}_{i b s} \\
&\hspace{10pt} + \f{J'}{2} \sum_{i,a\neq b,s\neq s'} c_{i a s}^{\dagger}
c_{i a s'}^{\dagger}c^{\phantom{\dagger}}_{i b s'}c^{\phantom{\dagger}}_{i b s}, 
\end{aligned}
\label{eq:Interaction}
\end{equation}
where $i$ is the lattice site, $a$ is the orbital index, and $n_{i a s} = c_{i a s}^{\dagger}c_{i a s}$ is the density operator. We assume that the phenomenological parameters satisfy $U' = U-2J$ and $J' = J$~\cite{DagottoEA11}. Since we will consider the weak-coupling limit in the following, this leaves the single parameter $J/U$ characterizing the interaction.

We base our analysis on the weak-coupling scheme for repulsive Hubbard models, introduced and developed in Refs.~\onlinecite{KohnLuttinger65, BaranovEA92, KaganChubukov89, ChubukovEA92, BaranovChubukovEA92, ChubukovEA93, HironoEA02, Hlubina99, RaghuEA10, Raghu2EA10, WeejeeEA14,ScaffidiEA14, SimkovicEA16, Scaffidi2017}. We first diagonalize $H_0$ and index the eigenstates by a band index $\mu=\alpha,\beta,\gamma$ and a pseudo-spin index, which we mostly keep implicit below. In this basis, the linearized gap equation reads
\begin{equation}
\sum_{\nu} \int_{S_{\nu}} \f{\D \bo{k}_{\nu}}{\lvert S_{\nu} \rvert} g(\bo{k}_{\mu}, \bo{k}_{\nu}) \varphi(\bo{k}_{\nu}) = \lambda \varphi(\bo{k}_{\mu}), 
\label{eq:GapEquation}
\end{equation}
where $S_{\nu}$ is the Fermi surface of band $\nu$, with $|S_{\nu}|$ the corresponding Fermi surface area, and $g$ the dimensionless matrix
\begin{equation}
g(\bo{k}_{\mu}, \bo{k}_{\nu}) = \sqrt{\f{\rho_{\mu} \bar{v}_{\mu}}{v_{\mu}(\bo{k}_{\mu}) } } \Gamma(\bo{k}_{\mu}, \bo{k}_{\nu}) \sqrt{\f{\rho_{\nu} \bar{v}_{\nu}}{v_{\nu}(\bo{k}_{\nu}) }  \vphantom{\f{\rho_{\mu} \bar{v}_{\mu}}{v_{\mu}(\bo{k}_{\mu}) }}  }.
\label{eq:gmatrix}
\end{equation}
Here, $\Gamma$ is the two-particle interaction vertex (see Ref.~\onlinecite{Scaffidi2017} for details) at leading (second) order, $\rho_{\mu} = \lvert S_{\mu} \rvert /[\bar{v}_{\mu} (2\pi)^3 ]$ is the density of states, and $\bar{v}_{\mu}^{-1} = \int_{S_{\mu}} \frac{\D \bo{k}}{\lvert S_{\mu} \rvert} v_{\mu}(\bo{k})^{-1}$.  Approaching the weak-coupling limit $U/t \to 0$ asymptotically, an eigenfunction $\varphi$ of the Eq.~\ref{eq:GapEquation} corresponding to a negative eigenvalue $\lambda$ yields the superconducting gap
\begin{equation}
\Delta(\bo{k}_{\mu}) \sim  \sqrt{\f{v_{\mu}(\bo{k}_{\mu})}{\bar{v}_{\mu}\rho_\mu}  } \varphi(\bo{k}_{\mu})
\label{eq:GapFunction}
\end{equation}
below the critical temperature $T_c \sim W \ e^{ -1/\lvert \lambda \rvert }$, where $W$ is the bare bandwidth. 

Since we have chosen a pseudo-spin basis which is consistent with the tetragonal point group, each eigenvector $\varphi$ belongs to one of its ten irreducible representations  $D_{4h}$~\cite{SigristUeda91, RaghuEA10, Annett1990}. In Table \ref{tab:Representations} we list all ten irreducible representations of the tetragonal point group, with corresponding order parameter structures, using the standard decomposition of the order parameter into even-parity, $d_0(-\bo{k}) = d_0(\bo{k})$, and odd-parity, $\bo{d}(-\bo{k}) = - \bo{d}(\bo{k})$, components as~\cite{BalianWerthamer63}
\begin{equation}
\Delta_{s s'}(\bo{k}) = \big[ \left( d_0(\bo{k}) \mathds{1} + \bo{d}(\bo{k})\cdot \bo{\sigma} \right) i\sigma_y \big]_{s s'}, 
\label{eq:DeltaDecomp}
\end{equation}
where $s$ and $s'$ are pseudo-spin indices. Only the odd-parity $E_u$ and even-parity $E_g$ representations are two-dimensional and permit TRSB order without the need of fine tuning model parameters. 
\begin{table}[h!tb]
\caption{Irreducible representations of the tetragonal point group  $D_{4h}$~\cite{SigristUeda91}. Even-parity representations (subscript $g$) are described by a scalar ($d_0$) order parameter, while odd-parity (subscript $u$) order parameters are described by a vector ($\bo{d}$); see Eq.~\eqref{eq:DeltaDecomp}. In the second column one should associate $f_j$ with any function that transforms like $\sin{k_j}$ under the point group operations, and $f_j^2$ with a function that transforms like $\cos{k_j}$ for $j = x, y, z$. Representations $E_u$ and $E_g$ are two-dimensional and can favor TRSB combinations as indicated.}
\begin{center}
\begin{tabular}{p{3.0cm} p{4.0cm} }
\toprule
 Representation & Order parameter  \\ \hline
 $A_{1g}$ & $d_0(\bo{k}) = f_x^2+f_y^2$  \\
 $A_{2g}$ & $d_0(\bo{k}) = f_x f_y (f_x^2-f_y^2)$ \\
 $B_{1g}$ & $d_0(\bo{k}) = f_x^2-f_y^2$  \\
 $B_{2g}$ & $d_0(\bo{k}) = f_x f_y$  \\
 $E_{g}$ & $d_0(\bo{k}) = f_z (f_y \pm i f_x) $ \\
 $A_{1u}$ & $\bo{d}(\bo{k}) = f_x \hat{x} + f_y \hat{y}$  \\
 $A_{2u}$ & $\bo{d}(\bo{k}) = f_y \hat{x} - f_x \hat{y}$  \\
 $B_{1u}$ & $\bo{d}(\bo{k}) = f_x \hat{x} - f_y \hat{y}$  \\
 $B_{2u}$ & $\bo{d}(\bo{k}) = f_y \hat{x} + f_x \hat{y}$  \\
 $E_{u}$ & $\bo{d}(\bo{k}) = (f_x \pm i f_y) \hat{z}$  \\ \toprule
\end{tabular}
\end{center}
\label{tab:Representations}
\end{table}

The Pauli principle assures that odd-parity (respectively, even parity) solutions correspond to pseudo-spin triplets (respectively, singlets). One should, however, keep in mind that a Zeeman field, as considered in Sec.~\ref{sec:MagneticSusceptibility}, couples to the physical spin, and not the pseudo-spin, which means that the behavior of the magnetic susceptibility cannot be deduced from the parity of the order parameter alone, and always requires a numerical calculation.

The method outlined above is valid in the limit of weak interactions. One might reasonably object that this limit is not strictly satisfied in the case of SRO (and possibly other real materials)~\cite{VaugierEA12, MravljeEA11, HuoEA13, BehrmannEA12}. However, recent work (Ref.~\onlinecite{Romer2EA19}) reviewing a range of weak- and strong-coupling numerical methods applied to the two-dimensional Hubbard model found that both the $\bo{k}$-dependency and the symmetry of the superconducting order exhibit surprisingly little variation between methods, suggesting a smooth transition from weak to strong coupling. Our aim is that the present study, using an advanced three-dimensional tight-binding model and a controlled approximation, provides helpful guidance for future studies.

\section{Results and discussion}
\label{sec:Results}
\begin{figure}[t]
	\centering
	\includegraphics[width=0.85\linewidth]{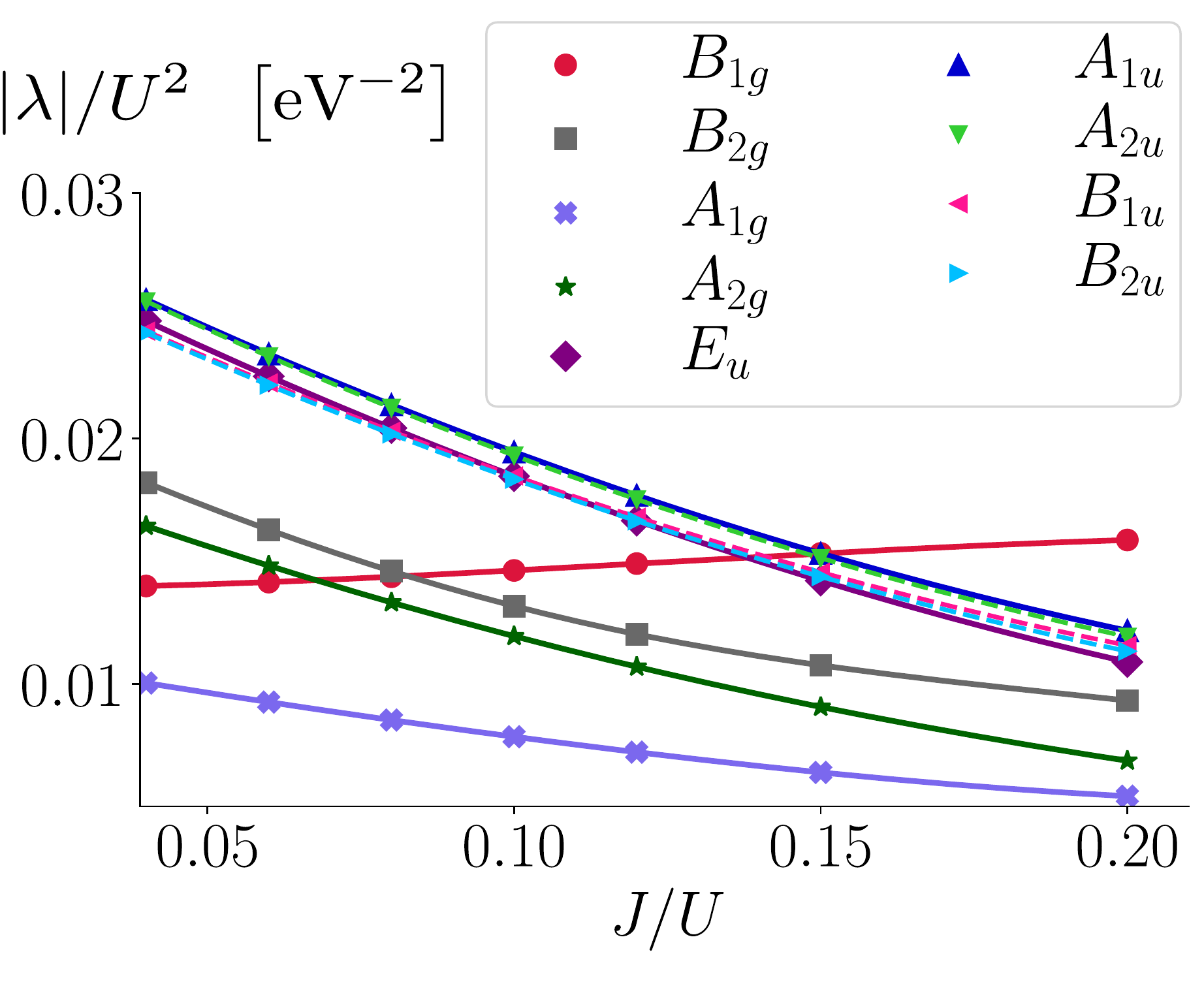}
	\caption{Eigenvalues of the leading order parameter in each irreducible representation: $A_{1g}$, extended $s$-wave singlet; $B_{1g}$, $B_{2g}$, $d$-wave singlet; $A_{2g}$, $g$-wave singlet; $E_u$, chiral $p$-wave triplet; and $A_{1u}$, $A_{2u}$, $B_{1u}$, $B_{2u}$, helical $p$-wave triplet.}
	\label{fig:CompareEigs}
\end{figure}

\begin{figure*}[h!tb]
	\centering
	\subfigure[]{\includegraphics[width=0.31\linewidth]{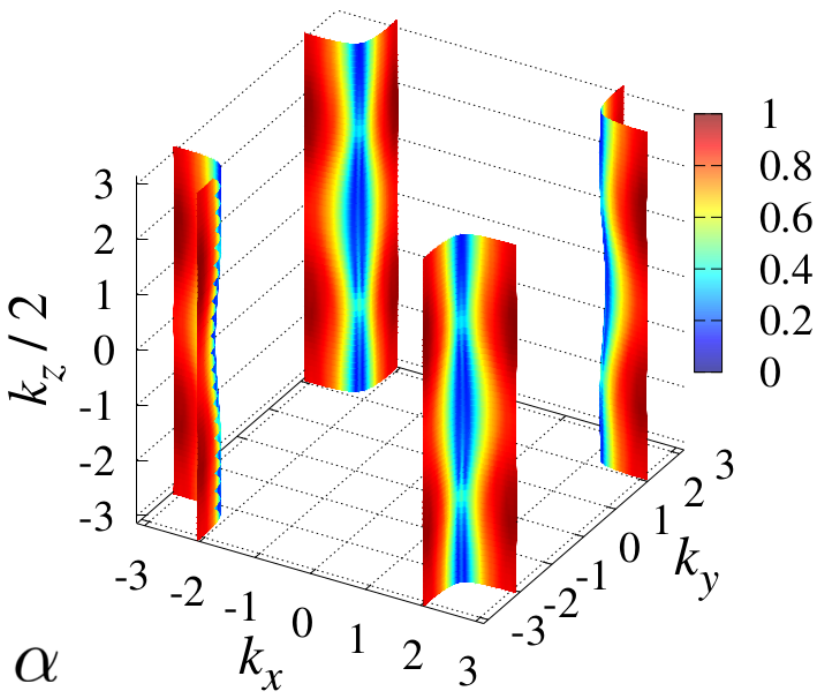}} \quad 
	\subfigure[]{\includegraphics[width=0.31\linewidth]{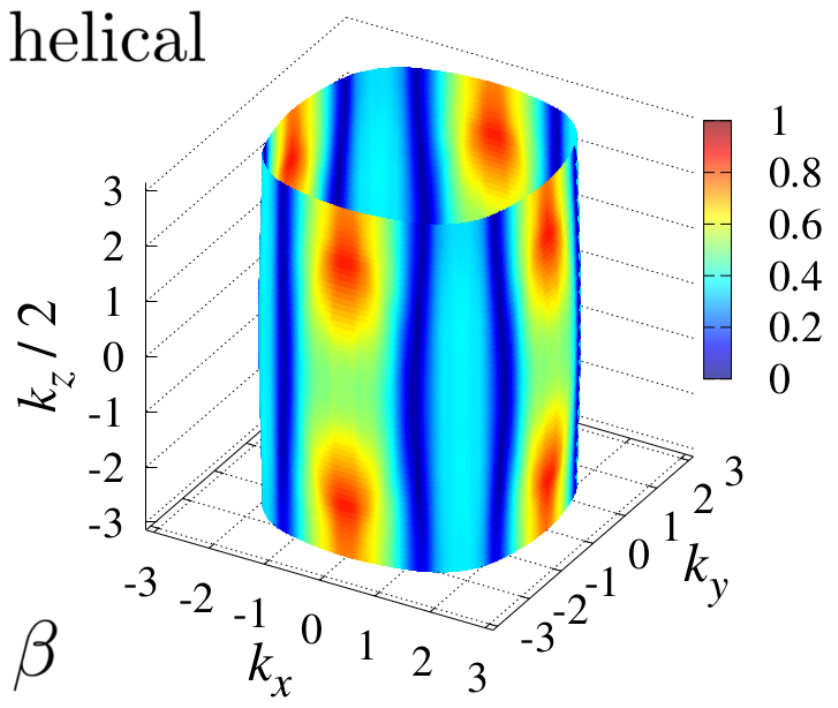}} \quad 
	\subfigure[]{\includegraphics[width=0.31\linewidth]{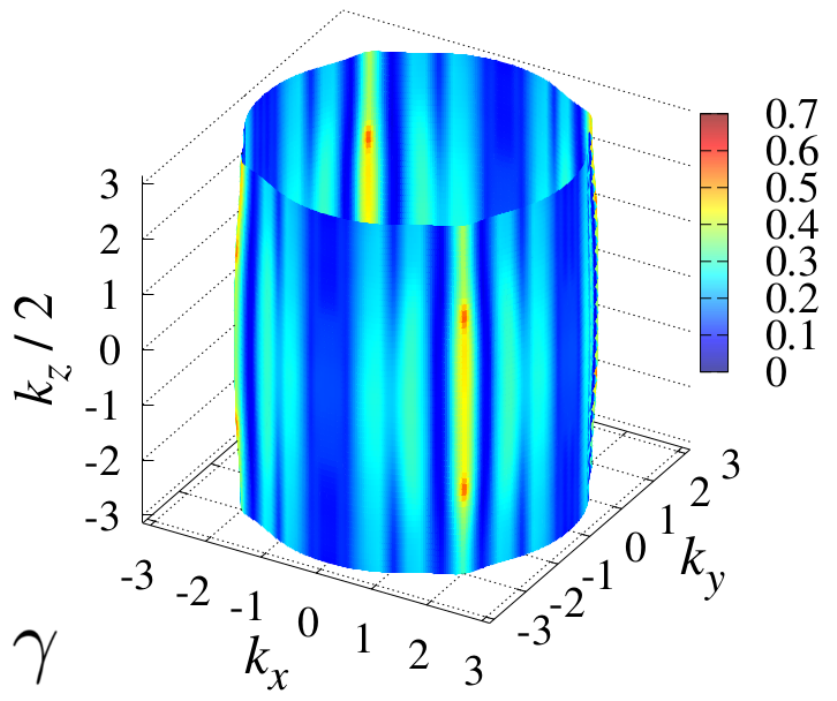}} \quad
    \subfigure[]{\includegraphics[width=0.31\linewidth]{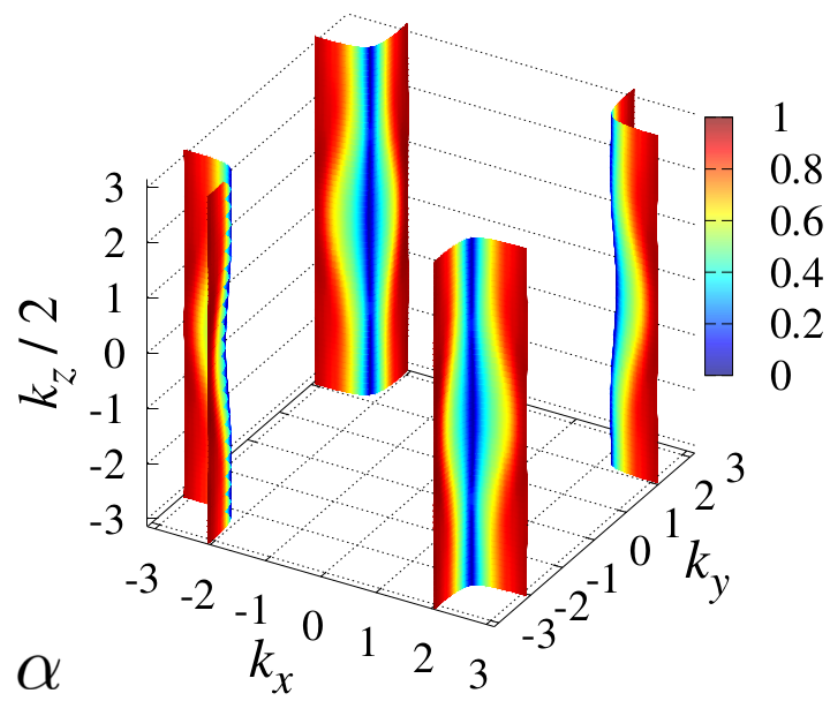}} \quad
	\subfigure[]{\includegraphics[width=0.31\linewidth]{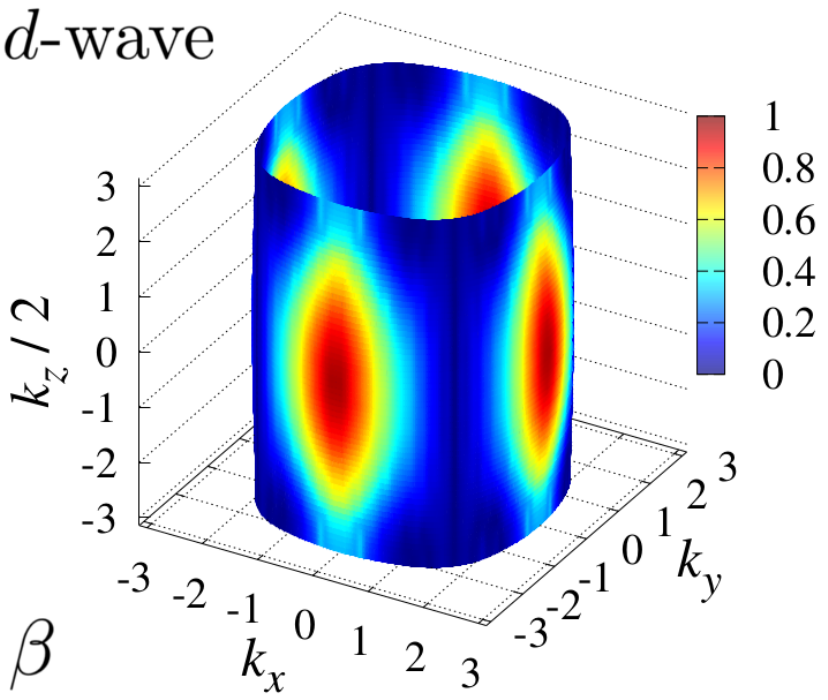}} \quad 
	\subfigure[]{\includegraphics[width=0.31\linewidth]{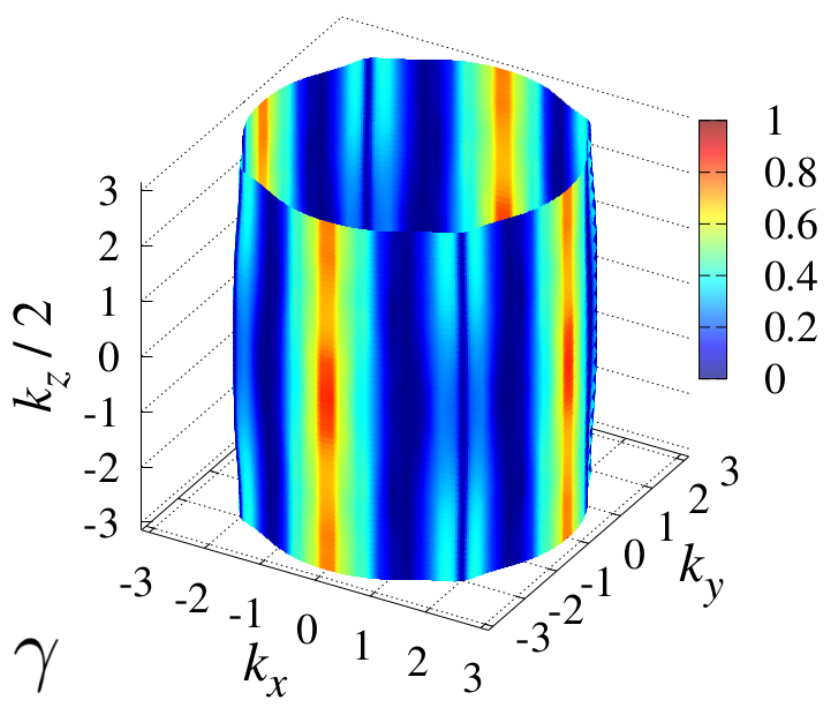}}
	\caption{(a) -- (c) $\lvert \Delta_{\mu} (\bo{k}) \rvert$ for the leading helical order parameter ($A_{1u}$) at $J/U = 0.06$, and (d) -- (f)  for the leading even-parity order parameter ($B_{1g}$) at $J/U = 0.20$. }
	\label{fig:LeadingGaps}
\end{figure*}

For a repulsive interaction, there is no superconducting instability at first order~\footnote{This can be proven by noticing that the (even-parity) vertex at first order can be recast as a \emph{Gramian} matrix.}. At second order, which we limit ourselves to in the calculation presented in the main text, the vertex we calculate is described in depth in Refs.~\onlinecite{Scaffidi2017, RaghuEA10, WeejeeEA14}. The leading eigenvalues in each irreducible representation are displayed as a function of $J/U$ in Fig.~\ref{fig:CompareEigs}. Whereas the even-parity orders show qualitatively different trends with $J/U$, the odd-parity states all show the same trend, and the splitting between them always remains small. The highest-$T_c$ state is the odd-parity helical order $A_{1u}$ for $J/U < 0.15$, and the even-parity $d_{x^2-y^2}$ order $B_{1g}$ for $J/U>0.15$. Both types of order parameter are shown in Figs.~\ref{fig:LeadingGaps}, with $k_z$ cuts displayed in Figs.~\ref{fig:SlicesHelical} and \ref{fig:SlicesB1g}. 

The results presented here were obtained by discretization of the Fermi surfaces, using $3318$ $\bo{k}$-points for each of the three Fermi sheets shown in Figs.~\ref{fig:FermiSurfaces} (a) -- \ref{fig:FermiSurfaces} (c) ($20$ points in the $k_z$ direction), thus solving the linearized gap Eq.~\eqref{eq:GapEquation} as a regular matrix eigenvalue problem. In the numerical implementation, the eigenvalues of Eq.~\eqref{eq:KineticHamiltonianMain} appear repeatedly in the two-particle vertex~\cite{Scaffidi2017} of Eq.~\eqref{eq:gmatrix}, and an effective diagonalization routine for the kinetic Hamiltonian constitutes the bulk of the numerical procedure~\cite{Kopp08}. Numerically, we computed sub-blocks of the $g$-matrix (Eq.~\eqref{eq:gmatrix}) simultaneously, using about $500$ cores over the duration of a few weeks.

\subsection{Gap structure}
\label{sec:GapStructure}
\begin{figure*}[h!tb]
	\centering
\subfigure[]{\includegraphics[width=0.31\linewidth]{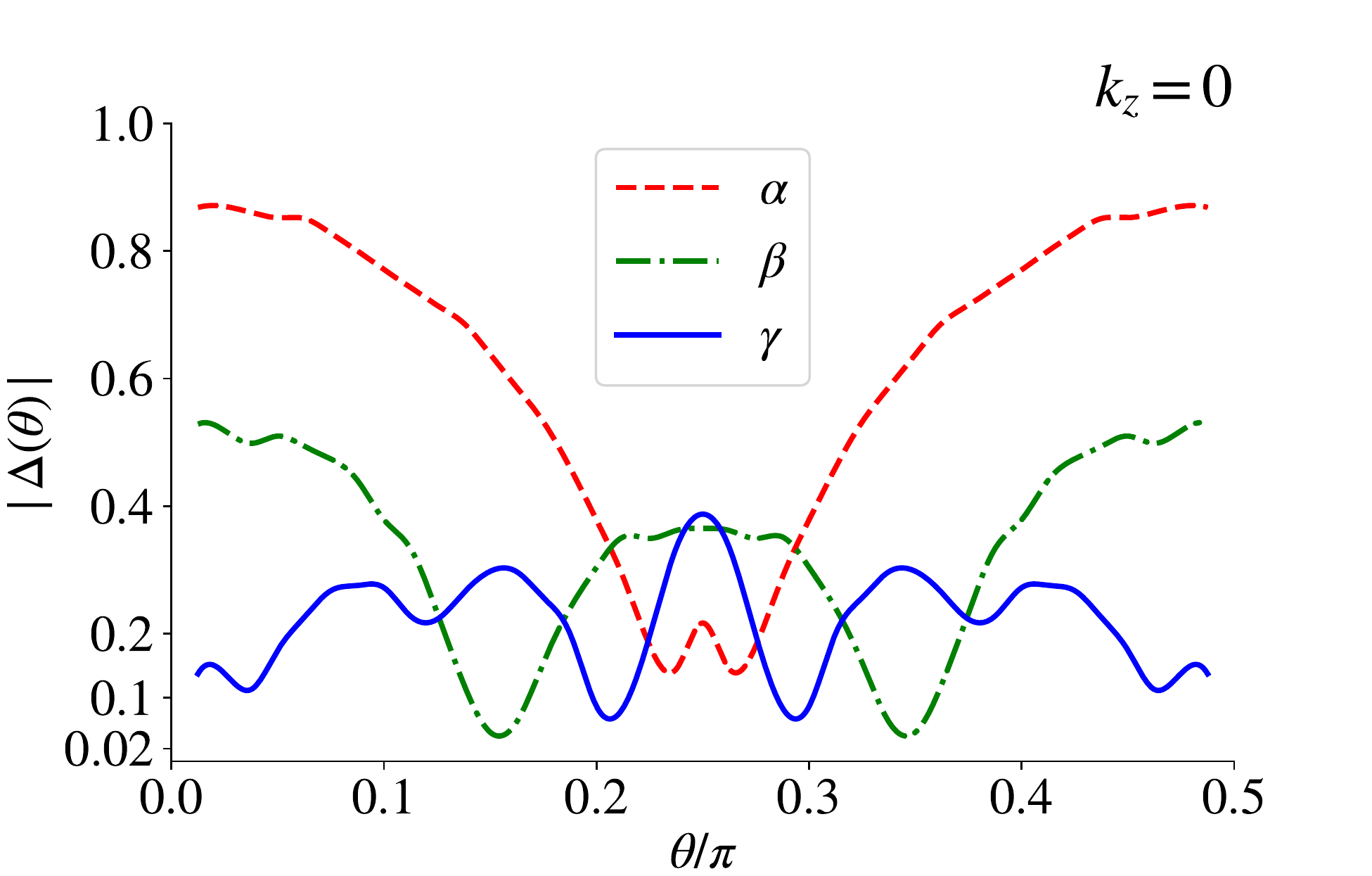}}\quad
	\subfigure[]{\includegraphics[width=0.31\linewidth]{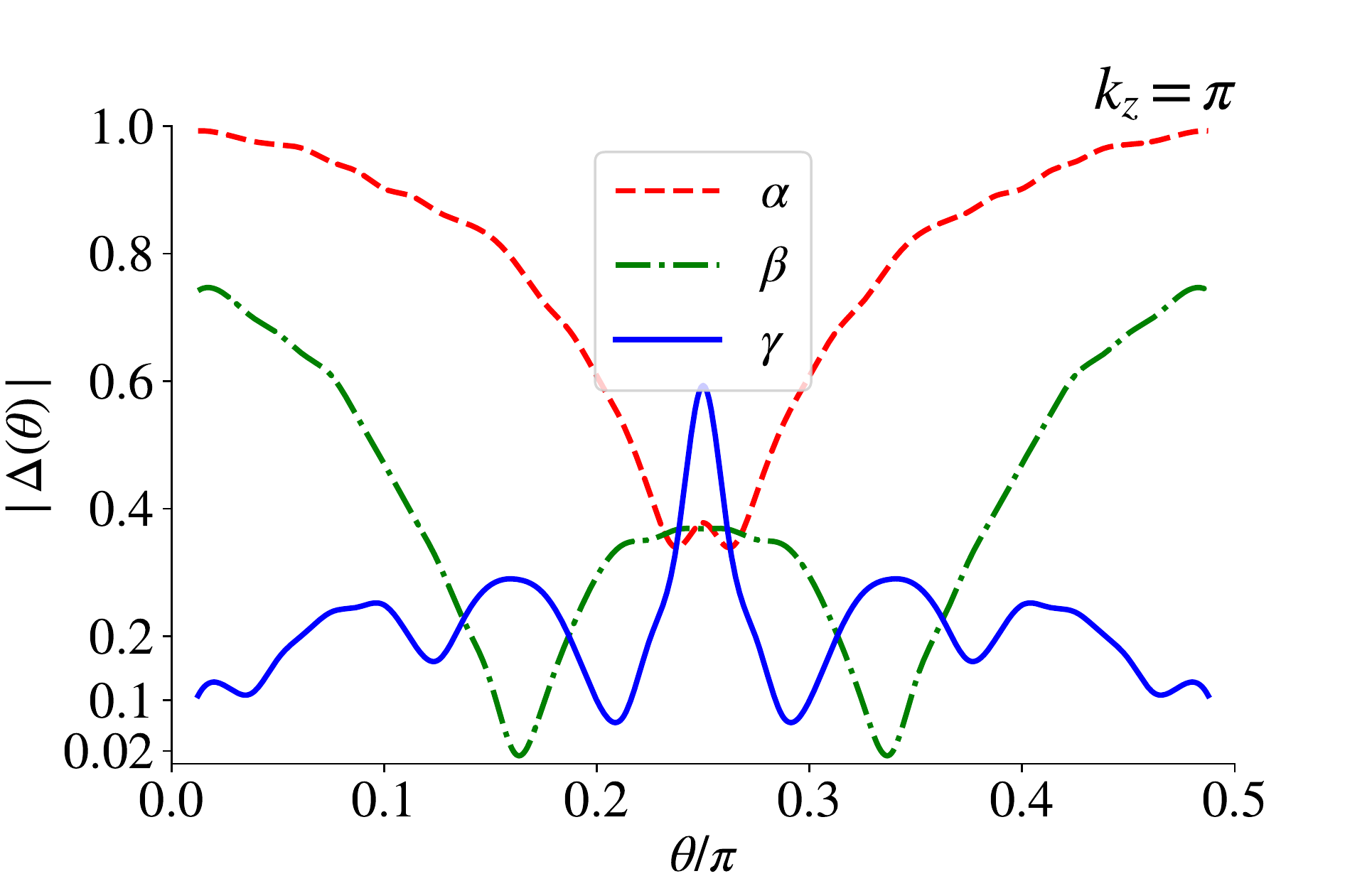}} \quad 
	\subfigure[]{\includegraphics[width=0.31\linewidth]{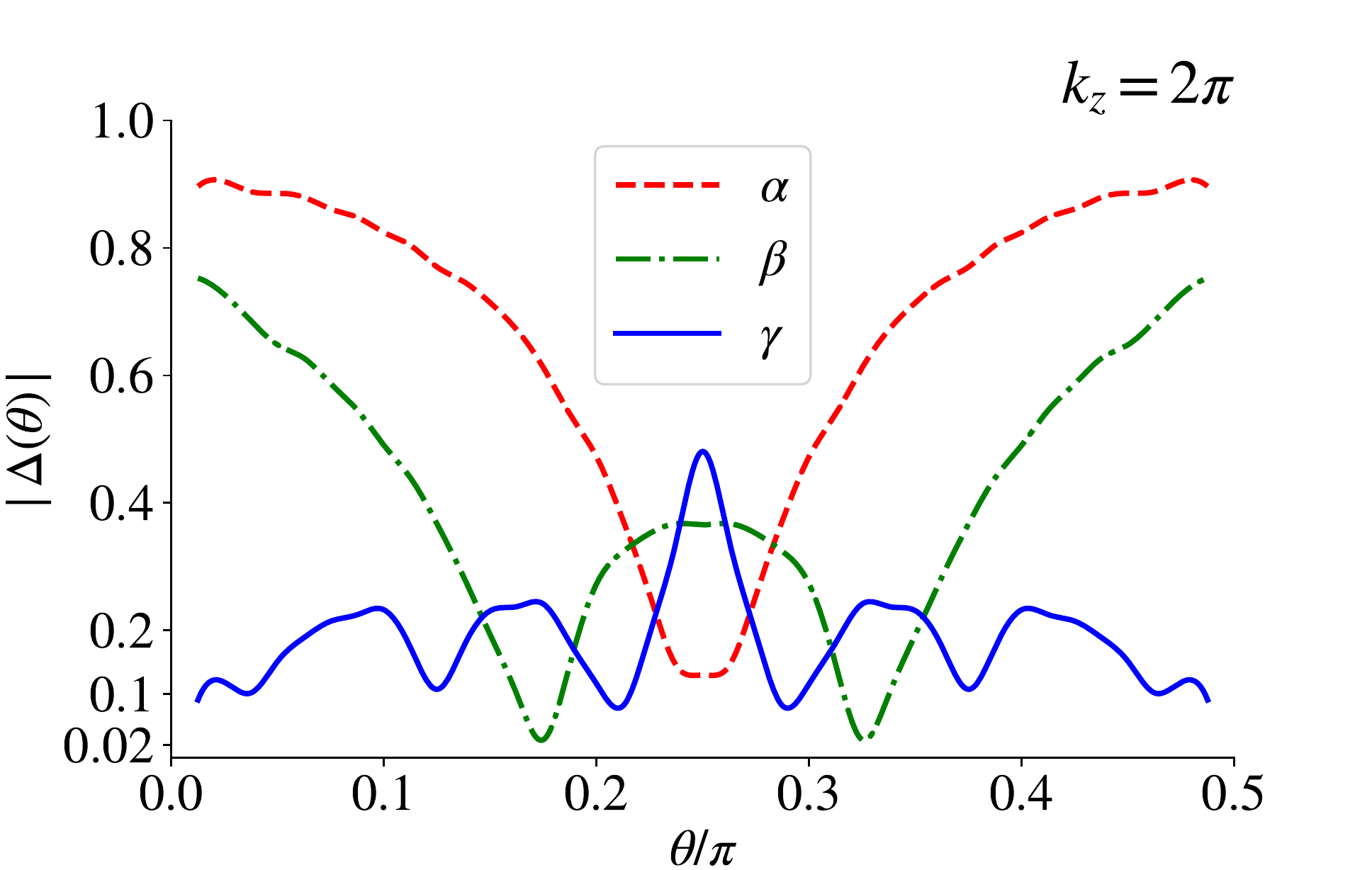}} 
\caption{Cuts for three values of $k_z$, showing the magnitude of the helical order parameter at $J/U = 0.06$. Here, $\theta$ is the in-plane polar angle, defined with vertex at $(0,0,k_z/2)$ for $\beta$ and $\gamma$, and vertex at $(\pi,\pi,k_z/2)$ for $\alpha$.}
	\label{fig:SlicesHelical}
\end{figure*}

\begin{figure*}[h!tb]
	\centering
	\subfigure[]{\includegraphics[width=0.31\linewidth]{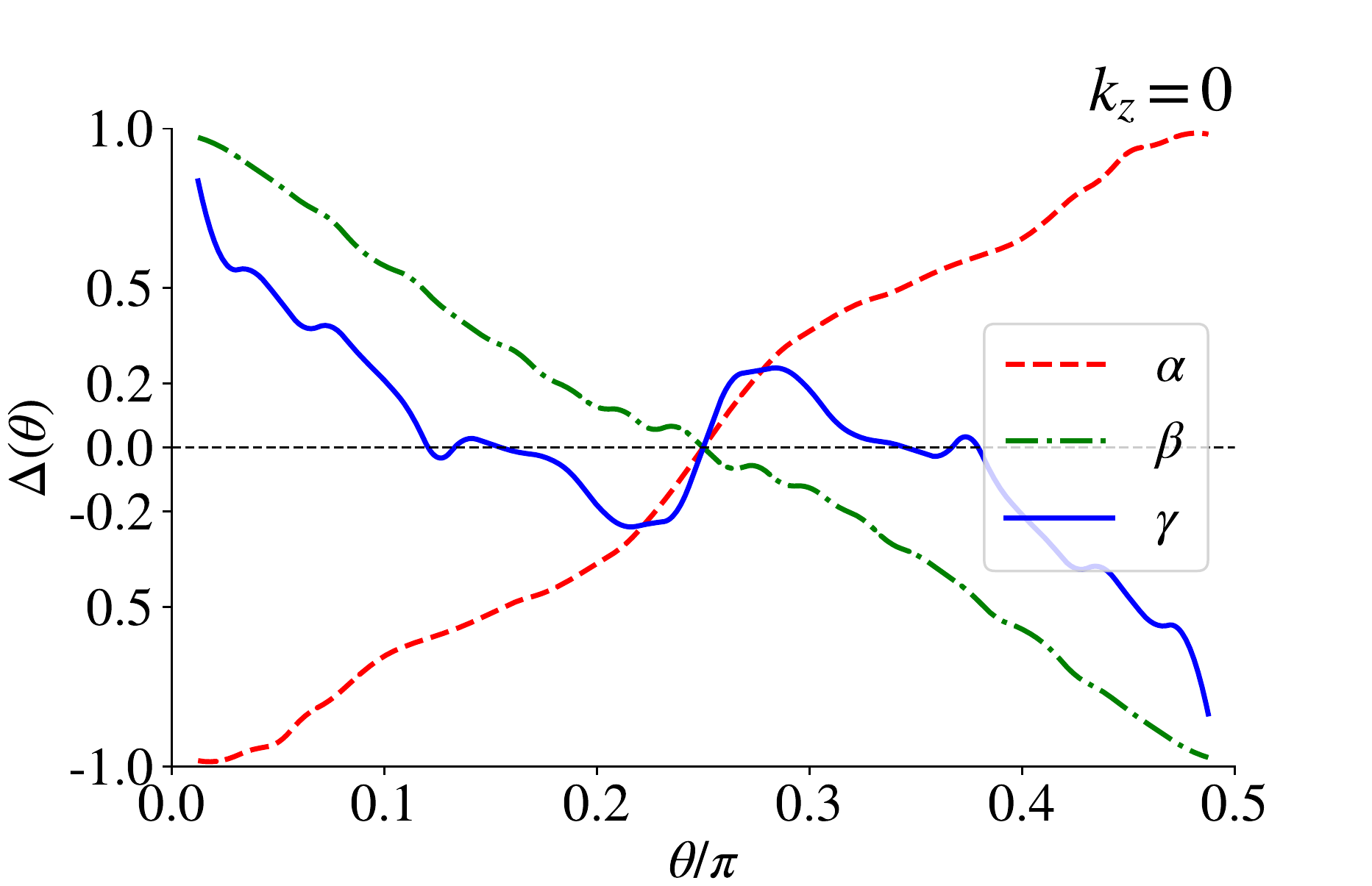}} \quad 
	\subfigure[]{\includegraphics[width=0.31\linewidth]{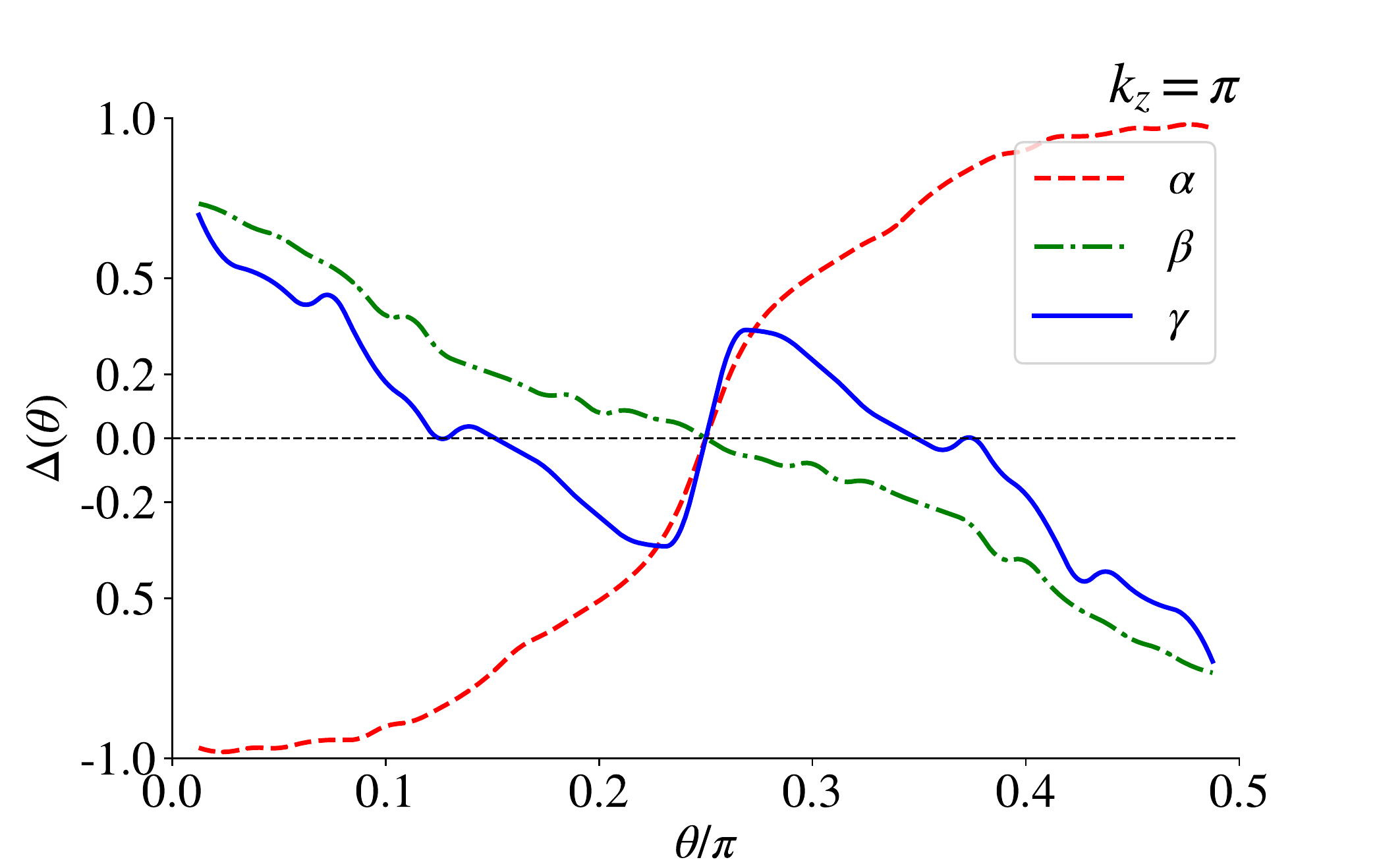}} \quad 
	\subfigure[]{\includegraphics[width=0.31\linewidth]{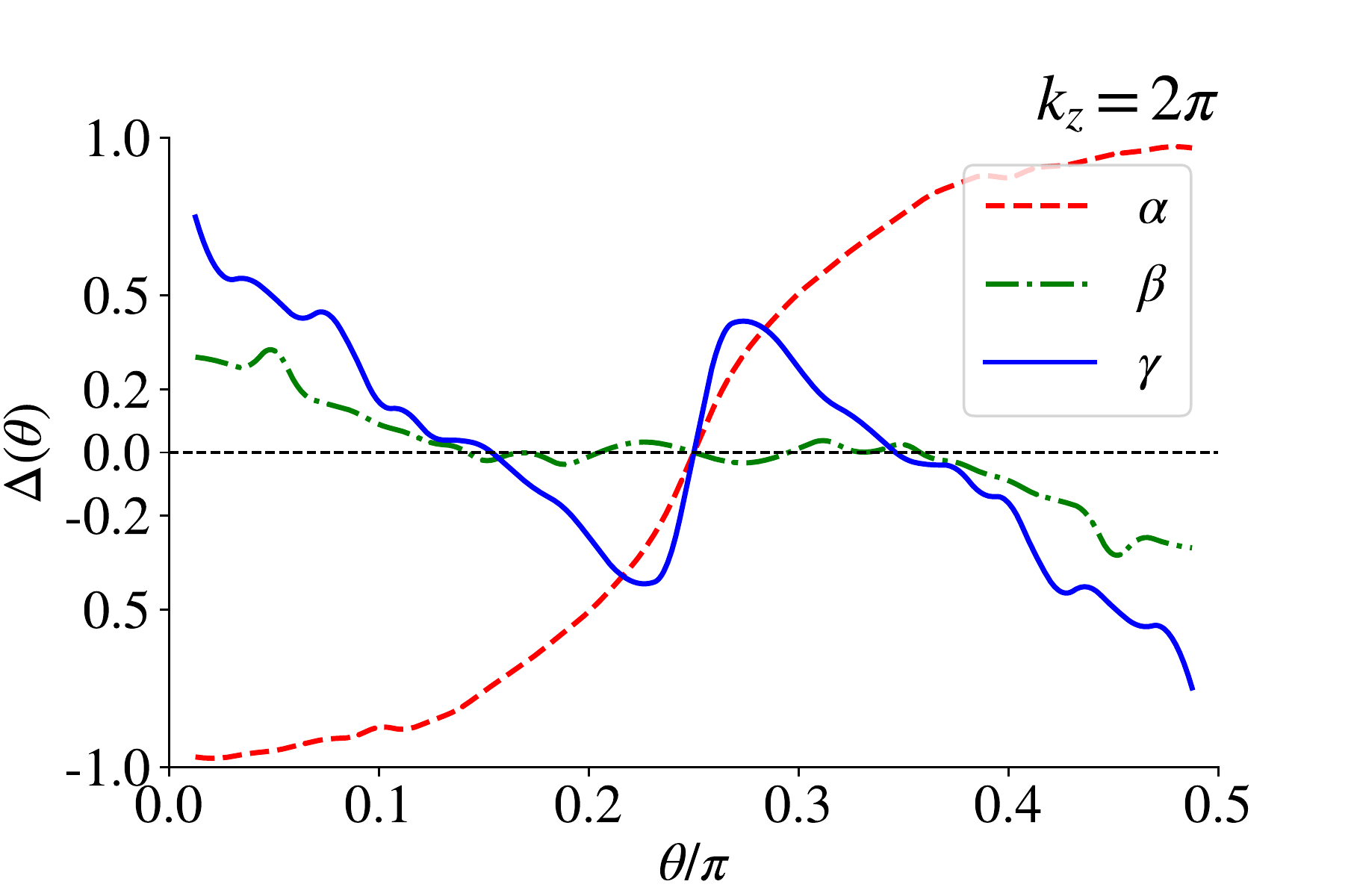}}
\caption{Same as described in the caption of Fig.~\ref{fig:SlicesHelical}, but here showing the signed $B_{1g}$ order parameter at $J/U = 0.20$.}
	\label{fig:SlicesB1g}
\end{figure*}

We focus first on the odd-parity helical phase realized for $J/U < 0.15$. The magnitude of the helical order parameter is displayed in Figs.~\ref{fig:LeadingGaps} (a) -- \ref{fig:LeadingGaps} (c). Deep vertical minima, as first predicted in two dimensions~\cite{ScaffidiEA14, ZhangEA18}, are present on all three bands. Notably, the gap on the $\beta$ band has $\min_{\theta} \lvert \Delta_{\beta}(\theta, k_z \approx \pi) \rvert \lesssim 0.02 \Delta_0$, where $\theta$ is the in-plane azimuthal angle and $\Delta_0$ the maximal gap, at locations in agreement with previous predictions~\cite{FirmoEA13, ScaffidiEA14, DodaroEA18}. The character of the minima also appear likely to agree with thermal Hall measurements~\cite{HassingerEA17, DodaroEA18}. For larger values of the Hund's coupling an even-parity $B_{1g}$ phase is realized. The order parameter, shown in Figs.~\ref{fig:LeadingGaps} (d) -- \ref{fig:LeadingGaps} (f), has symmetry-imposed vertical line nodes on all bands and additionally a suppressed gap in large regions of the $\beta$ and $\gamma$ sheets. We note that in two dimensions the $B_{1g}$ phase does not appear in the weak-coupling limit for the physical range of $J/U$, but it does appear with the random phase approximation for finite values of $U$~\cite{ScaffidiEA14, ZhangEA18, RomerEA19}. In Figs.~\ref{fig:SlicesHelical} and \ref{fig:SlicesB1g} we display $k_z$ cuts of the helical and $d$-wave order parameters at $J/U = 0.06$ and $J/U = 0.20$, respectively. Note in particular the (near-)nodes on $\beta$ in Fig.~\ref{fig:SlicesHelical} (b).

In three dimensions, the possibility of an $E_g$ order parameter with a horizontal line node at $k_z = 0$ emerges~\cite{ZuticMazin05}. Interest in this state has been fueled by recent specific heat measurements~\cite{KittakaEA18} combined with the possibility of explaining both TRSB and a nodal gap. However, this sector turns out to be strongly disfavored in our weak-coupling limit: At (\emph{e.g.}) $J/U = 0.20$, the best candidate has $\lambda_{E_g}/\lambda_{B_{1g}} \approx 0.03$ and thus does not come close to competing with the semi-two-dimensional order parameters found.

\subsection{Specific heat}
\label{sec:SpecificHeat}
The eigenvector calculated at weak coupling, $\varphi_{\mu}(\bo{k})$, is related to the order parameter via Eq.~\eqref{eq:GapFunction}. We assume further that the order parameter factorizes as $\Delta(T) \Delta_{\mu}(\bo{k})$, with $\max_{\bo{k}} \Delta_{\mu}(\bo{k}) = 1$. The temperature dependency of $\Delta(T)$ is assumed to be that of a conventional BCS superconductor~\cite{Tinkham75}. The generalized BCS relation~\cite{Sigrist05}, fixing the overall size of the gap, when given the experimental value of $T_c \approx 1.48$~K, is
\begin{equation}
\f{\Delta(0)}{k_B T_c} = \pi \exp(- \gamma - \llangle \log{\lvert \Delta(\bo{k}) \rvert} \rrangle_{\mathrm{FS}} ),
\label{eq:BCSrelationgeneral}
\end{equation}
where $\gamma \approx 0.5772$ is Euler's constant, and where we introduced the average
\begin{equation}
\llangle \log{\lvert \Delta(\bo{k}) \rvert} \rrangle_{\mathrm{FS}} = \f{\sum_{\mu} \int_{S_{\mu}} \D \bo{k} \hspace{1mm} \f{\lvert \Delta_{\mu}(\bo{k}) \rvert^2 }{v_{\mu}(\bo{k})}  \log{\lvert \Delta_{\mu}(\bo{k})\rvert}}{\sum_{\nu} \int_{S_{\nu}} \D \bo{k}  \hspace{1mm} \f{\lvert \Delta_{\nu}(\bo{k}) \rvert^2 }{v_{\nu}(\bo{k})} }.
\label{eq:average}
\end{equation} 
For a uniform gap, the average in Eq.~\eqref{eq:average} is 0, and the BCS relation $\Delta(0) = 1.764~k_B T_c$ is recovered~\cite{Tinkham75}. For a multiband superconductor the specific heat per temperature per normal state value, $\gamma_n$, can be expressed as~\cite{Sigrist05}
\begin{widetext}
\begin{equation}
\frac{C(T)}{T \gamma_n} = \frac{3}{2\pi^2(k_B T)^3} \int_0^{\infty} \D \xi \hspace{1mm} \Big\langle \f{\xi^2 + \Delta(T)^2 \lvert \Delta_{\mu}(\bo{k}) \rvert^2 - \f{T}{2} \f{\partial \Delta(T)^2}{\partial T}  \lvert \Delta_{\mu}(\bo{k}) \rvert^2}{\cosh^2(\f{E_{\mu}(\bo{k})}{2k_B T})} \Big\rangle_{\mathrm{FS}},
\label{eq:specificheat}
\end{equation}
\end{widetext}
where $E_{\mu}(\bo{k}) = \sqrt{ \xi^2 + \Delta(T)^2  \lvert \Delta_{\mu}(\bo{k}) \rvert^2}$, and where the average here is evaluated as 
\begin{equation}
\langle A \rangle_{\mathrm{FS}} = \f{1}{\sum_{\nu} \rho_{\nu}}  \sum_{\mu} \f{\rho_{\mu}}{\lvert S_{\mu} \rvert} \int_{S_{\mu}} \D \bo{k} \hspace{1mm} A.
\label{eq:average2}
\end{equation}

\begin{figure}[t]
	\centering
	\subfigure[]{\includegraphics[width=\linewidth]{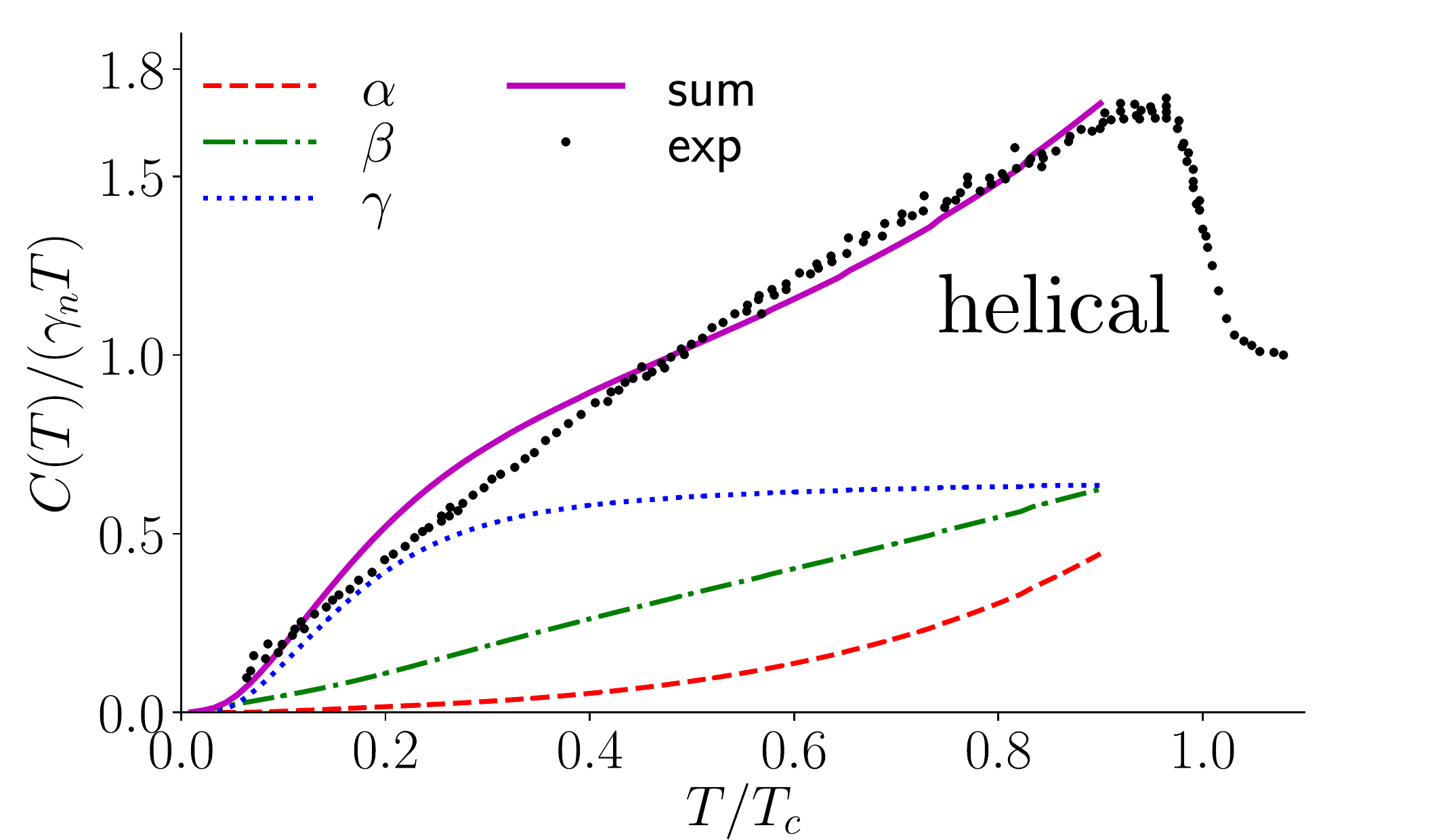} } \quad 
	\subfigure[]{\includegraphics[width=\linewidth]{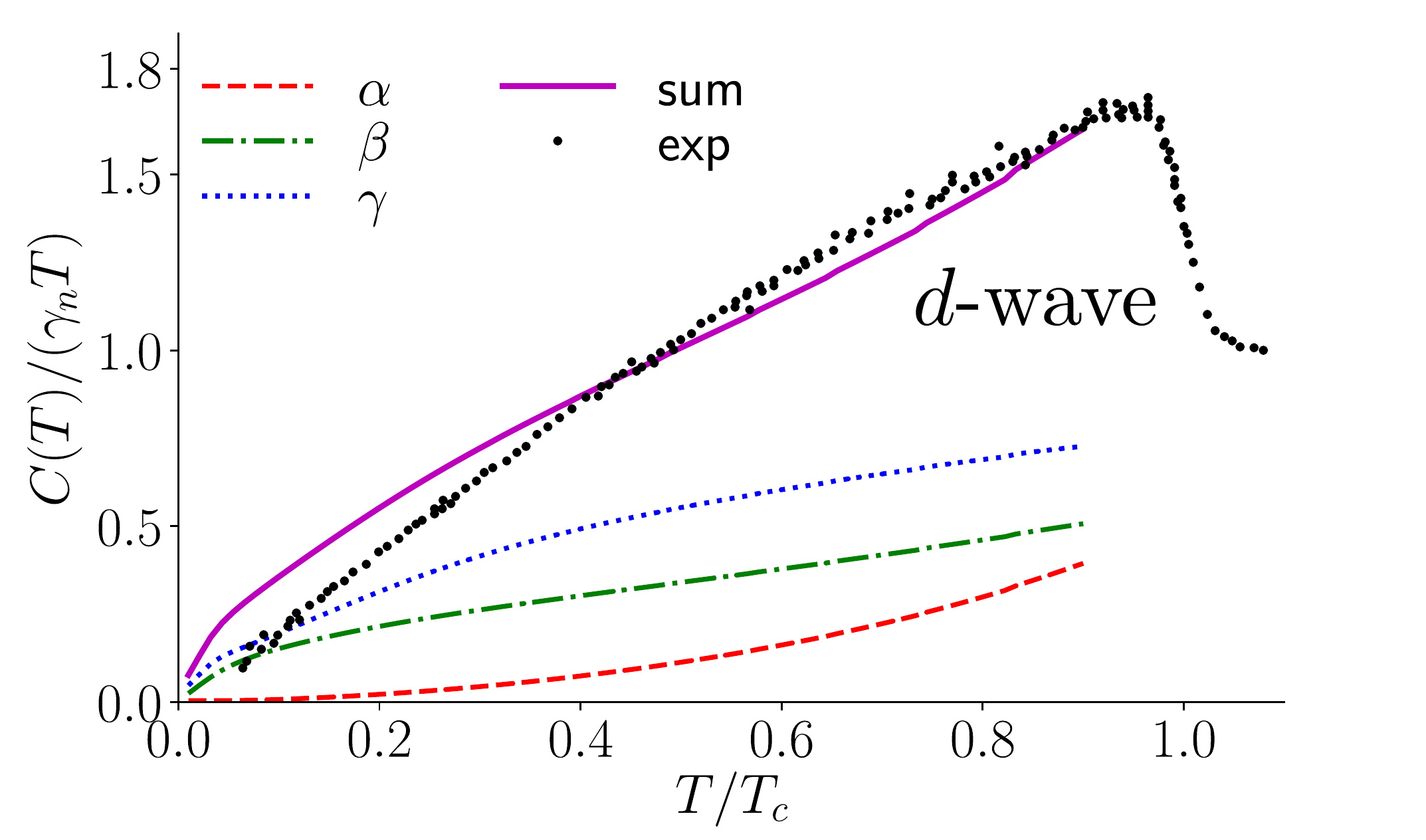} }  
	\caption{Specific heat $C$ divided by temperature $T$ and normal state value $\gamma_n$, calculated for leading (a) helical triplet order with $J/U = 0.06$, and (b) $d$-wave singlet order with $J/U = 0.20$. The black dots are experimental values adapted from Ref.~\onlinecite{NishiZakiEA00}.}
	\label{fig:SpecificHeat}
\end{figure}
In Fig.~\ref{fig:SpecificHeat} we show the specific heat as computed from Eq.~\ref{eq:specificheat} compared to experimental values from Ref.~\onlinecite{NishiZakiEA00}. For the helical order parameter (at $J/U = 0.06$) obtained at weak coupling, the minima are deep enough to practically behave as accidental vertical nodes, and to provide a decent agreement with specific heat data, as shown in Fig.~\ref{fig:SpecificHeat} (a). The $B_{1g}$ order parameter (at $J/U = 0.20$) also provides a fairly good match with specific heat data, as revealed by Fig.~\ref{fig:SpecificHeat} (b). In contrast to previous studies~\cite{NishizakiEA98}, there is no fine tuning of model parameters in these calculations (the specific heat dependency on $J/U$ is weak). Crucially, and as pointed out in previous work, a gap comparable in size on all three bands seems necessary to reproduce the features of the experimental data~\cite{NishizakiEA98, MackenzieEA17, ScaffidiEA14}.

\subsection{Magnetic susceptibility}
\label{sec:MagneticSusceptibility}
Since both leading order parameters exhibit nodal behavior in practice, we  turn to a different probe of the superconducting order: the spin susceptibility, as measured by the NMR Knight shift~\cite{PustogowEA19, IshidaEA19}. 

We consider the normal state Hamiltonian with Zeeman terms when an external magnetic field is applied,
\begin{equation}
H = H_0 + H_S + H_L,
\label{eq:H}
\end{equation}
where $H_0$ is here the normal state Hamiltonian in the absence of a magnetic field (Eq.~\eqref{eq:Hamiltonian}), $H_S$ ($H_L$) denotes coupling between the magnetic field $\bo{H}$ and the spin $\bo{S}$ (orbital $\bo{L}$) degrees of freedom (cf.~Ref.~\onlinecite{RamiresEA17}),
\begin{align}
H_0 &= \sum_{\bo{k}, s, s'} \sum_{a,b,c} \left( \varepsilon_{a b}(\bo{k}) +i \eta \epsilon_{a b c} \sigma^c_{s s'} \right) c_{a s}^{\dagger}(\bo{k}) c_{b s'}(\bo{k}), \label{eq:H0} \\
H_S &= - 2 \mu_B \bo{H} \cdot \bo{S} \nonumber \\
&= - \mu_B \sum_{\bo{k}, s, s'} \sum_{a, b} H_b \sigma^b_{s s'}  c_{a s}^{\dagger}(\bo{k}) c_{a s'}(\bo{k}), \label{eq:HS} \\
H_L &= - \mu_B \bo{H} \cdot \bo{L} \nonumber \\ 
&= - \mu_B i \sum_{\bo{k}, s, s'} \sum_{a,b,c} H_c \epsilon_{a b c} c_{a s}^{\dagger}(\bo{k}) c_{b s'}(\bo{k}). \label{eq:HL}
\end{align}
Here, $\varepsilon_{a b}(\bo{k})$ are the single-particle orbital terms, $\sigma^a$ is the $a$'th Pauli matrix, $\mu_B$ is the Bohr magneton (the vacuum permeability was fixed to $\mu_0 =1$), $\epsilon_{a b c}$ is the Levi-Civita symbol, and $a,b,c$ run over orbitals $xz, yz, xy$ ($A, B, C$, respectively).

The matrix $U(\bo{k})$ is defined to diagonalize the matrix $\pazocal{H}_{0}(\bo{k}) + \pazocal{H}_{S} + \pazocal{H}_{L} $, where 
\begin{align}
H &= \sum_{\bo{k}} \bo{\psi}^{\dagger}(\bo{k}) \left( \pazocal{H}_{0}(\bo{k}) + \pazocal{H}_{S} + \pazocal{H}_{L} \right) \bo{\psi}(\bo{k}), \label{eq:MagneticHamiltonian} \\
\bo{\psi}(\bo{k}) &= [c_{A \uparrow}(\bo{k}), c_{B  \uparrow}(\bo{k}), c_{C  \downarrow}(\bo{k}), c_{A \downarrow}(\bo{k}), c_{B  \downarrow}(\bo{k}), c_{C \uparrow}(\bo{k})]^T. \label{eq:StateVector}
\end{align}
The (six) Fermi surfaces are defined by $\xi_{\mu  \sigma}(\bo{k}) = 0$ for $\mu = \alpha, \beta, \gamma $ and $\sigma = +, - $. In the absence of a magnetic field the two energies $\xi_{\mu  \pm}(\bo{k})$ become degenerate due to restored time-reversal symmetry. The matrix $U(\bo{k})$ and the energies $\xi_{\mu \sigma}(\bo{k})$ contain all information needed to calculate the magnetization in the normal state, as defined below. The transformation between orbitals/spins ($a$, $s$) and bands/pseudo-spins ($\mu$, $\sigma$) is given by the components of $U(\bo{k})$:
\begin{equation}
c_{a s}(\bo{k}) = \sum_{\mu, \sigma} [u^{\mu  \sigma}_{a  s}(\bo{k})]^{\ast} c_{\mu  \sigma}(\bo{k}).
\label{eq:Transformations}
\end{equation}
We define the magnetization (with the magnetic field and the measured response along direction $i = x, y, z$) as
\begin{align}
\pazocal{M}^{ii} &\equiv \f{1}{2} \sum_{a, s_1, s_2 } \sigma_{s_1 s_2}^i M_{a}^{s_1 s_2} \big\rvert_{\bo{H} \parallel \hat{i}},
\label{eq:MagnetizationTensor} \\
M_{a}^{s_1 s_2} &= \mu_B \sum_{\bo{k}} \langle c_{a s_1}^{\dagger}(\bo{k}) c_{a s_2}(\bo{k}) \rangle.
\label{eq:Magnetization}
\end{align}
Using Eq.~\eqref{eq:Transformations} and $\langle c_{\mu \sigma_1}^{\dagger}(\bo{k}) c_{\nu \sigma_2}(\bo{k}) \rangle = \delta_{\mu  \nu} \delta_{\sigma_1 \sigma_2} f(\xi_{\mu \sigma_1}(\bo{k}))$, where $f$ is the Fermi function, the matrix elements of Eq.~\eqref{eq:Magnetization} are given by
\begin{equation}
M_{a}^{s_1 s_2} = \mu_B \sum_{\bo{k}, \mu, \sigma} u^{\mu \sigma}_{a s_1}(\bo{k}) \left[ u^{\mu \sigma}_{a s_2}(\bo{k}) \right]^{\ast}  f(\xi_{\mu \sigma}(\bo{k})).
\label{eq:MagnetizationNormal}
\end{equation}
In the superconducting phase we add superconducting terms at orbital level
\begin{equation}
H_{\Delta} = \sum_{\bo{k}} \sum_{a_1, a_2, s, s'} \Delta_{s_1  s_2}^{a_1 a_2}(\bo{k}) c_{a_1 s}^{\dagger}(\bo{k}) c_{a_2 s'}^{\dagger}(-\bo{k}) + \text{h.c.}, 
\label{eq:HDelta}
\end{equation}
\begin{equation}
\Delta_{s_1 s_2}^{a_1 a_2}(\bo{k}) = \big[ \left( d_0^{a_1 a_2}(\bo{k}) \mathds{1} + \bo{d}^{a_1 a_2}(\bo{k})\cdot \bo{\sigma} \right) i\sigma_y \big]_{s_1 s_2}. 
\label{eq:DeltaOrb}
\end{equation}
The electron operators in orbital and spin basis are now expressed as linear combinations of their particle ($u$'s) and hole ($v$'s) constituents,
\begin{equation}
c_{a s} (\bo{k}) = \sum_{\mu, \sigma} \left( u^{\mu \sigma}_{a s}(\bo{k})  c_{\mu  \sigma}(\bo{k}) + v^{\mu \sigma}_{a  s}(-\bo{k})  c^{\dagger}_{\mu  \sigma}(-\bo{k}) \right).
\label{eq:PHSdecomposition}
\end{equation}
In turn, this leads to the magnetization matrix elements
\begin{widetext}
\begin{equation}
M_{a}^{s_1 s_2} = \mu_B \sum_{\bo{k}, \mu, \sigma}  \left( \big[ u^{\mu \sigma}_{ a s_1}(\bo{k}) \big]^{\ast} u^{\mu \sigma}_{a s_2}(\bo{k}) f(E_{\mu  \sigma}(\bo{k})) + \big[ v^{\mu \sigma}_{a s_1}(-\bo{k}) \big]^{\ast} v^{\mu \sigma}_{a s_2}(-\bo{k})  [1-f(E_{\mu \sigma}(-\bo{k}))] \right).
\label{eq:MagnetizationSC}
\end{equation}
\end{widetext}
Note that due to particle-hole symmetry the two terms of Eq.~\eqref{eq:MagnetizationSC} contribute equally. To relate the orbital gaps of Eq.~\eqref{eq:DeltaOrb} to the order parameters obtained at weak coupling, we make use of the transformation
\begin{equation}
\begin{aligned}
\Delta_{s_1 s_2}^{a_1 a_2}(\bo{k}) = \sum_{\mu, \sigma_1, \sigma_2}  &\Delta_{\sigma_1  \sigma_2}^{\mu}(\bo{k}) \big[ u_{a_1 s_1}^{\mu \sigma_1}(\bo{k})u_{a_2 s_2}^{\mu \sigma_2}(-\bo{k}) \\
& + u_{a_1 s_1}^{\mu \sigma_2}(\bo{k})u_{a_2 s_2}^{\mu \sigma_1}(-\bo{k}) \big],
\end{aligned}
\label{eq:TransformationToOrbitalBasis}
\end{equation}
where $\Delta_{\sigma_1  \sigma_2}^{\mu}(\bo{k})$ is the order parameter in band and pseudospin basis, and where $u_{a s}^{\mu \sigma}(\bo{k})$ are eigenvector components of $\pazocal{H}_0(\bo{k})$ (\emph{i.e.}~in the absence of a magnetic field, crucially with the same gauge choice as in the weak-coupling calculation). 

We use the spin susceptibility normalized by its normal state value as a proxy for the Knight shift in the superconducting state,
\begin{equation}
K^i(T) = \Delta\pazocal{M}^{ii}(T)/\Delta\pazocal{M}^{ii}_{n} = \chi^{ii}(T)/\chi^{ii}_n,
\label{eq:KnightShift}
\end{equation}
where the subscripts refer to the normal state value, $\Delta$ indicates a small increment in the external magnetic field in the linear response regime, and $\chi^{ii}$ is a diagonal element of the magnetic susceptibility tensor. In the numerical evaluation of the momentum integral of Eq.~\eqref{eq:MagnetizationSC} we associate for any given $\bo{k}$ the weak-coupling order parameter solution from the Fermi surface point closest to $\bo{k}$, convoluted with a Gaussian damping factor set by the distance from the Fermi surface. In practice, the orbital coupling in Eq.~\eqref{eq:HL} did not change the resulting Knight shift by any significant amount and was consequentially not included in the numerical results presented here.

\begin{figure}[h!tb]
	\centering
	\subfigure[]{\includegraphics[width=\linewidth]{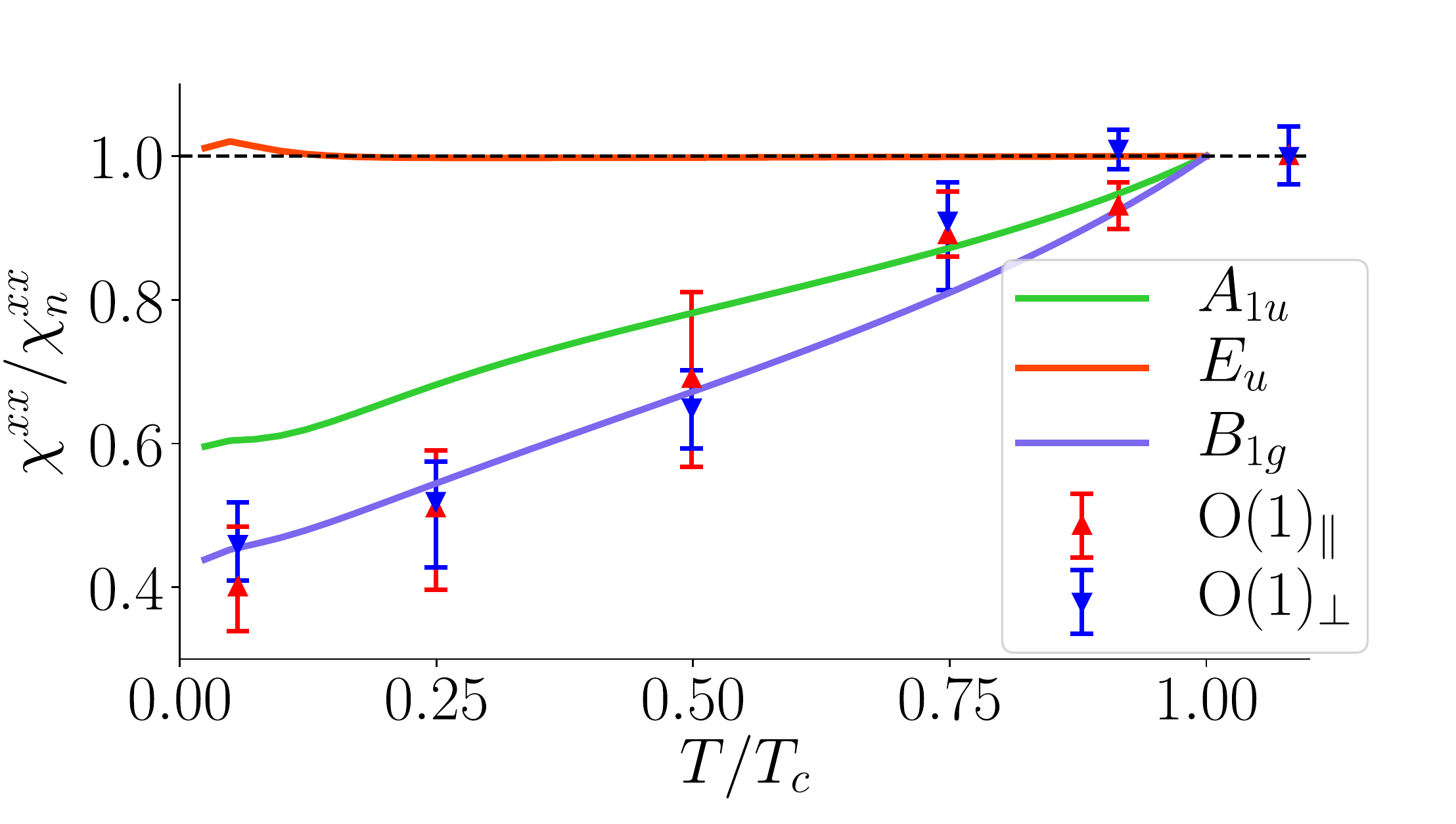}} \quad 
	\subfigure[]{\includegraphics[width=\linewidth]{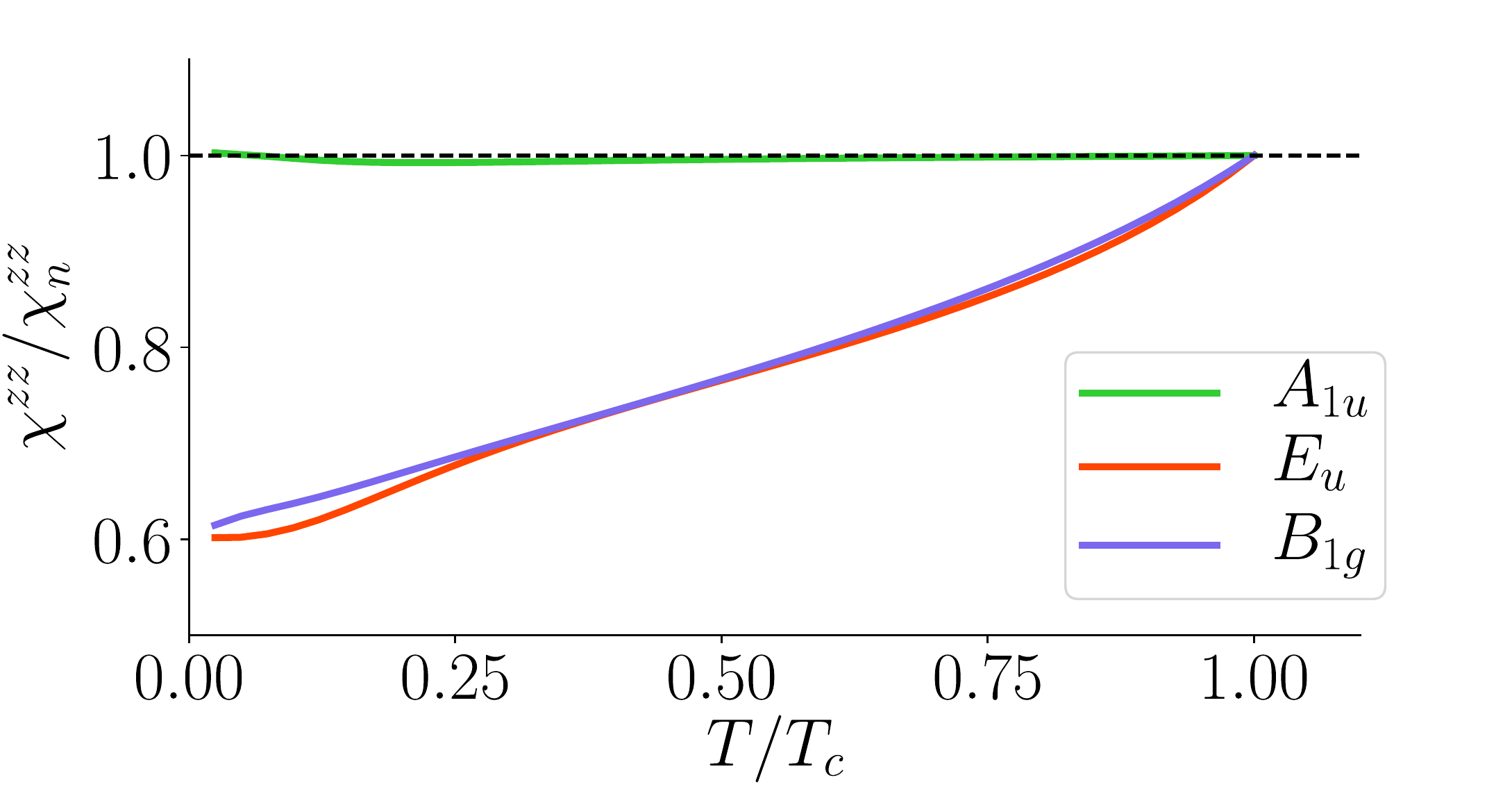}} \quad 
	\caption{The temperature dependence of the calculated magnetic susceptibilty (normalized by the normal state value) for three relevant order parameters, $A_{1u}$ and $E_u$ orders at $J/U = 0.06$, and $B_{1g}$ order at $J/U = 0.20$. (a) External field pointing in the basal plane. NMR Knight shift data from Ref.~\onlinecite{IshidaEA19} for two different Oxygen sites are plotted along with the calculated magnetic susceptibility. (b) External field pointing out of the basal plane. }
	\label{fig:NMR_temperature}
\end{figure}

With SOC, the normal state spin susceptibility has a `bulk' interband contribution that is not related to the Fermi surface, and that is therefore not affected by superconductivity. Further, as mentioned before, Cooper pairs only form well-defined singlets (respectively, triplets) in the pseudo-spin basis, but not in the physical spin basis. This leads to similar values of $K^x$ for both order parameters at zero temperature: $K^x(T=0)=0.45$ for the $d$-wave order at $J/U=0.20$ and $K^x(T=0)=0.59$ for the helical order at $J/U=0.06$. These numbers are in rough agreement with a recent NMR experiment~\cite{PustogowEA19}, which indicates a drop of around $K^x(T=20~\text{mK}) \approx 0.5$.  The only order parameter clearly seen to conflict with the experimental value is the chiral $p$-wave, which shows almost no drop ($K^x(T=0)=0.99$). Note that, for textbook order parameters without SOC, there would have been a sharp contrast between $d$-wave ($K^x(T=0)=0$) and helical ($K^x(T=0)=0.5$)~\cite{RiceSigrist95}. 

Recently, the experimental results of Ref.~\onlinecite{PustogowEA19} were reproduced by Ref.~\onlinecite{IshidaEA19}, where the temperature dependence of the Knight shift was also measured. Values of $K^x(T)$ and $K^z(T)$ for representative order parameters are presented in Fig.~\ref{fig:NMR_temperature}. With the the current experimental data it appears difficult to sharply distinguish a helical from a $d$-wave order parameter. However, the low-temperature values of $K^{x}$ arguably appear in best agreement with $d$-wave order. Crucially, it would be desirable, although perhaps technically difficult, to have the NMR experiment repeated for out-of-plane fields as this would yield a crisp way to distinguish helical from $d$-wave order, as Fig.~\ref{fig:NMR_temperature} (b) shows.

\section{Conclusions}
\label{sec:Conclusions}
Both the $d$-wave and helical orders found in this calculation have vertical (near-)nodes, and seem compatible with specific heat data and recent Knight shift measurements~\cite{PustogowEA19}. However, despite fairly strong SOC, a chiral order parameter appears incompatible with the observed Knight shift drop.  Further microscopic multiband Knight shift calculations would help in quantifying this, and an out-of-plane NMR experiment would help in definitely distinguishing the helical states from even-parity order parameters. While the OPs exhibit a substantial $k_z$ dependence on the $\beta$ band, they remain overall fairly two-dimensional. We do not see any microscopic evidence of a favored $E_g$ gap with symmetry-imposed horizontal line nodes, at least in the weak-coupling limit.

An important outstanding aspect requiring further assessment, both theoretically and experimentally, is how to unify evidence of TRSB with either helical or even-parity order~\cite{SatoshiEA19}. Should it turn out that TRSB is spurious, or unrelated to superconductivity, the scenario of a single-component $d$-wave order parameter would become a natural contender. Another possibility would be the formation of a two-component order parameter which couples different irreducible representations with accidentally close critical temperatures~\cite{HuangEA18, HuangNematicEA19,RomerEA19}. The near-degeneracy of the various odd-parity orders, found here and in previous work~\cite{ScaffidiEA14, ZhangEA18}, could potentially provide evidence for this scenario. We note that the four helical states in general are non-degenerate when $J/U \neq 0$, $\eta \neq 0$, and $\varepsilon_{AB} \neq 0$~\cite{ScaffidiEA14}. However, the splitting of these states is a relatively small effect, as Fig.~\ref{fig:CompareEigs} shows. The (accidental) formation of a TRSB combination of helical states would be non-unitary. 

%
\section*{Acknowledgments}
%
Helpful conversations with Fabian Jerzembeck, Clifford Hicks, Steven Kivelson, Yoshiteru Maeno, Daniel Agterberg, Assa Auerbach, Stephen Blundell, Mats Horsdal, Andrew Mackenzie, Peter Hirschfeld, and Catherine Kallin are acknowledged. We thank Kenji Ishida for providing the experimental data from Ref.~\onlinecite{IshidaEA19}. T.S. acknowledges support	from	the Emergent	Phenomena in Quantum	Systems	initiative	of	 the	Gordon	and	Betty	Moore	Foundation.	H.S.R.$\,$ and G.F.L.$\,$ are both supported by the Aker Scholarship.  F.F.$\,$acknowledges support from the Astor Junior Research Fellowship of New College, Oxford. S.H.S.$\,$is supported by EPSRC Grant No.~EP/N01930X/1.

\bibliography{ReferencesSRO}

\end{document}